\documentclass[pra,aps,twocolumn,amssymb,nofootinbib,showpacs%
amsmath]{revtex4-1}
\usepackage{hyperref}
\usepackage{epsfig}
\begin{document}
\title{Electromagnetic vacuum of complex media II: the Lamb shift and the total vacuum energy}
\author{M.~Donaire$^{1}$}
\email{mdonaire@fc.up.pt}
\affiliation{$^{1}$Centro de F\'{\i}sica do Porto, Faculdade de Ci\^{e}ncias da Universidade do Porto, Rua do Campo Alegre 687, 4169-007 Porto, Portugal}

\date{10 September 2011}

\begin{abstract}
We study the physical content of the  electromagnetic vacuum energy of a random medium made of atomic electric dipoles. First, we evaluate the contribution of statistical fluctuations to the average total vacuum energy, which is made out of the integration of the variations of the Lamb shift with respect to the coupling constant. While the Lamb shift is a function of the electrical susceptibility only, the vacuum energy is generally not. Second, we make clear why the effective medium bulk energy does not account for the total vacuum energy of a molecular dielectric. Consequently, the Lamb shift does not derive from the effective medium bulk energy except at leading order in the molecular density. The local field factors provide natural cutoffs for the spectrum of the total vacuum energy at a wavelength of the order of the correlation length. Third, we investigate to what extent shifts in the spectrum of the dielectric constant may be attributed to the binding energy of a dielectric. In particular, in the continuum approximation we have found a relation between the electrostatic binding energy and the Lorentz-Lorenz shift. Nonetheless, we conclude that the knowledge of the spectrum of the refractive index is insufficient either to quantify the energy of radiative modes or to estimate the electrostatic binding energy of molecular clusters.
\end{abstract}
 \maketitle

\section{Introduction}\label{1}
It has long since been recognized that the Casimir effects, the van der Waals (vdW) forces and the Lamb shift share a common origin \cite{Dzy,MilonniScripta1,Milonnibook}. It is customary to ascribe the vdW forces and the Lamb shift to the short-range interactions between the microscopic constituents of dielectric media. In contrast, the Casimir effects are attributed to the long-range interactions between macroscopic objects. Also, some authors refer to the energy of discrete modes as Casimir  energy while they term bulk energy that of the continuous spectrum \cite{Milonnietal,Molina,Milton}. In this paper we deal with a translation-invariant molecular dielectric made of atomic electric dipoles. Therefore, aside possible geometrical resonances in clusters, the spectrum of electromagnetic (EM) fluctuations is continuous and the above distinctions are unnecessary.\\
\indent We aim to compute the EM energy of the ground state of the dielectric. It has been also recognized that this energy can be interpreted in two alternative ways. That is, when speaking of the interaction energy of fluctuating dipoles it is referred to as binding energy \cite{Abriko,Obada,FelderhofvdW}. Alternatively, when computed out of the variations of the zero-point vacuum fluctuations of the EM field it is termed vacuum energy \cite{Boyer,Power}. Evidently, this distinction is just semantic.\\
\indent The EM energy manifests in the shifts of bound state levels and in the shifts of the spectrum of the dielectric constant.
However, there does not exist an obvious correspondence between them. The binding energies of the medium prior to the coupling of
the dipoles to radiation are those of the quantum atomic states and those of the short range forces which structure the spatial configuration of the dipoles.
The former can be parameterized by the bare resonant frequencies of the atomic transitions. The latter depend on the precise form of the interaction
potential between the molecules and reflect on the spatial correlation functions which give rise to the spatial dispersion of the dielectric function.
We will restrict our analysis to situations in which the correlation functions do not depend on dipole-dipole interactions. Therefore, the energy
shifts will affect only the atomic levels, even though they will depend on external degrees of freedom (d.o.f.). For this reason we will
refer to the energy shifts as generalized Lamb shifts. They are made of two contribution. The first one is common to all the dipoles and is
due to the coupling of each isolated dipole to bare radiation. That is the free-space Lamb shift \cite{Wylie,Power,Welton}, which is additive and
whose integration is a self-energy termed free space Lamb energy. The second contribution is due to the interaction of each dipole with the rest, which is mediated by the multiple scattering of virtual photons. That is the scattering Lamb shift \cite{Wylie,WelschI,WelschII}.
Physically, it is the energy needed (or released) in the removal of one of the dipoles out of the dielectric.
The vdW forces and torques acting upon each dipole derive from it \cite{PuffyMuhoraidAmJ,WelschII}.
Its integration is a non-additive binding energy which involves a number of dipoles and collective d.o.f. The non-additive character of
the scattering Lamb shift can be better explained in terms of the decay of excited atoms. When one of the dipoles is in an excited state,
its frequency is renormalized by a 'resonant' Lamb shift which contains a scattering contribution \cite{Wylie,WelschI,WelschII}. Therefore,
part of the energy released in its decay belongs to collective d.o.f. For this reason, when several neighboring dipoles are excited and decay
at a time the energy released is not the sum of their individual energy shifts.\\
\indent On the other hand, according to Schwinger's source theory, the vacuum energy can be computed out of the integration of the variations of an effective action for the EM field w.r.t. some effective coupling to the dielectric medium \cite{Schwinger2105,Schwinger4095,SchwingerMath}. Since the variations are taken adiabatic, the vacuum energy only depends on the initial and final properties of the dielectric. However, the result depends on both the action varied and the parameter of the variations.\\
\indent  In principle, there is no obvious correspondence between the shifts on the dispersion relations of optical modes and the Lamb shifts. To this respect, Feynman \cite{Feynman}, Power \cite{Power} and Milonni \cite{MilonniScripta2} have proved that the free-space Lamb shift can be computed from the variation of the vacuum energy due to the presence of a single atomic dipole. That result has been extended by Schaden, Spruch and Zhou \cite{Schaden} to the case of a uniform distribution of non-mutually interacting dipoles. In reference \cite{Milonnietal} Milonni, Schaden and Spruch have extrapolated those approaches to the computation of the Lamb shift from variations on the effective medium bulk energy of a molecular dielectric due to a change in the refractive index. However, that  result was not totally conclusive since near field effects and local field factors (LFFs) were neglected in the calculation. In fact, the findings in \cite{PRAvc} suggest that that result does not hold when proper account of LFFs and inherent correlations is taken.\\
\indent The paper is organized as follows. In Sec. \ref{S2} we define fundamental concepts and review the basic formulas obtained in \cite{PRAvc}. In Sec. \ref{S3} we explain the role of statistical fluctuations in the computation of the vacuum energy of a random medium. In Sec. \ref{S4} we explain why Schwinger's approach is not suitable to study the Lamb shift. In Sec. \ref{S5} we compute the Lamb shift and the vacuum energy density up to order two in the molecular density for a hard sphere model. The results are compared with the approach of Milonni \emph{et al.} \cite{Milonnietal} in Sec.\ref{discy}.  Sec. \ref{Continuum} deals with the EM binding energy of a molecular dielectric in the continuum approximation. We analyze previous approaches \cite{PRAvc,Obada} and comment on the possibility to find signatures of the vacuum energy on the spectrum of the dielectric constant. The conclusions are summarized in Sec. \ref{la6}.\\
\indent Regarding notation, we will label three-spatial-component vectors with arrows and three-by-three tensors with overlines. We will denote the Fourier-transform of functionals with $\vec{q},\omega$-dependent arguments instead of the $\vec{r},t$-dependent arguments of their space-time representation. Quantum operators will be denoted with bold letters and hats on top.

\section{Ground states, polarizabilities and propagators}\label{S2}
A  generic statistically homogeneous molecular dielectric is made of $N$ equivalent point electric dipoles in a volume $\mathcal{V}$, with a typical correlation length $\xi$, such that, in the  limit $\mathcal{V}\gg V\gg\xi^{3}$, the average numerical density is uniform, $V^{-1}\int_{V}\textrm{d}^{3}r\rho(\vec{r})=\rho=N/\mathcal{V}$. For the sake of simplicity we will consider the canonical ensemble at temperature $T$ with $\mathbf{Z}_{N}(T)=\int\prod_{i=1}^{N}\textrm{d}^{3}R^{i}\exp{[-U(\vec{R}^{1},..,\vec{R}^{N})/k_{B}T]}$ the $N$-body canonical partition function. $U$ is the interaction energy, which is considered a function of the external d.o.f. of the dipoles only. That is, of the center of mass positions and velocities, $\vec{R}_{i},\vec{v}_{i}$, $i=1,..,N$. Eventually, in the limit $k_{B}T\ll m_{i}v_{i}^{2},U^{max}$, $m_{i}$ being the scatterer mass and $U^{max}$ the potential barrier from which steric forces derive, we can neglect the scatterer dynamics and assume that the potential $U$ is short-ranged.
This way, we can ignore Doppler effects and guarantee that the correlation functions do not depend on the long-range dipole-dipole interaction which is the object of our study \cite{BulloughJPhysA2,Abriko}. The dielectric constant and rest of effective optical parameters are stochastic functions which admit cluster expansions. That is, series of $n$-scattering terms computed out of the convolution of $n$-body spatial correlation functions with single particle polarizabilities and electric field propagators --see below and \cite{BulloughHynne}. Further, we will work  for simplicity in the zero-temperature limit, $T=0$. In addition to the aforementioned inequalities it demands $k_{B}T\ll\hbar\omega_{0}$, with $\hbar\omega_{0}$ the typical excitation energy of the atomic dipoles, so that all the dipoles remain in their ground state.\\
\indent The interaction Hamiltonian in the electric dipole approximation reduces to
\begin{equation}\label{Hamil}
\hat{H}_{int}(t)=\int \textrm{d}^{3}r-e\hat{\vec{\mathbf{r}}}(\vec{r},t)\cdot\hat{\vec{\mathbf{E}}}(\vec{r},t),
\end{equation}
where $e$ is the electronic charge an $e\hat{\vec{\mathbf{r}}}(\vec{r},t)$, $\hat{\vec{\mathbf{E}}}(\vec{r},t)$ are the dipole moment density and electric field operators respectively in the Heissemberg representation. Before statistical averages being performed, we consider a fixed configuration of $N$ indistinguishable two-level atomic dipoles for which $-e\hat{\vec{\mathbf{r}}}(\vec{r},t)=\sum_{i=1}^{N}-e\hat{\vec{\mathbf{r}}}_{i}(t)\delta^{(3)}(\vec{r}-\vec{R}^{i})$,
with $e\hat{\vec{\mathbf{r}}}_{i}(t)$ the dipole moment operator associated to the dipole with position vector $\vec{R}^{i}$. The EM radiation is intended as a reservoir of infinite number
of d.o.f. while the internal d.o.f. of the dipoles constitute small systems. The coupling between the radiation reservoir and the dipoles
is weak so that ordinary time-dependent perturbative calculations can be carried out. In the following we give some definitions and explain the notation.
\subsection{EM vacuum and atomic ground states}
Prior to turning on the interaction Hamiltonian,
we denote the translation-invariant EM vacuum at $T=0$ by $|\Omega_{0}\rangle$ and the bare ground and excited atomic states, common to all the dipoles, by $|A_{b}^{i}\rangle$ and  $|B_{b}^{i}\rangle$, $i=1,..,N$, respectively. Dipole and EM fluctuations are uncoupled and uncorrelated.\\
\indent Next, consider the dipoles infinitely separated from each other or otherwise a unique dipole in free space, say the one at $\vec{R}^{i}$, and turn on the interaction Hamiltonian $-e\hat{\vec{\mathbf{r}}}_{i}(t)\cdot\hat{\vec{\mathbf{E}}}(\vec{R}^{i},t)$. In this case the EM vacuum  gets polarized locally only due to the dipole fluctuations. Reciprocally, the atomic states get renormalized by the EM fluctuations and radiation reaction, being denoted by $|A^{i}_{0}\rangle$, $|B^{i}_{0}\rangle$, which are common to all the dipoles.\\
\indent Further, when the $N$ dipoles are brought together to form a specific configuration, say $m$, with position vectors $\{\vec{R}^{i}_{m}\}$, $i=1,..,N$, the EM vacuum gets polarized  by the local and non-local dipole fluctuations and the atomic states get renormalized by both the local interaction of each dipole with the 'bare' EM fluctuations and the non-local mutual interactions between the fluctuating dipoles. Alternatively, it can be interpreted that each atomic state gets renormalized by the local coupling of the corresponding dipole moment to the polarized fluctuations of the EM field --i.e. polaritons. In this case the polarized vacuum is denoted by $|\Omega_{m}\rangle$, which is not translation invariant, and the renormalized atomic states are denoted by $|A^{i}_{m}\rangle$, $|B^{i}_{m}\rangle$, $i=1,..,N$, being all different in general. The states $|\Omega_{m}\rangle,|A^{i}_{m}\rangle,|B^{i}_{m}\rangle$ will be defined later on by their physical content in a way analogous to that in \cite{PRAvc,BulloughObadaPhysicaA}. That is, by the fluctuations of the EM field and dipole moment operators.\\
\indent Finally, consider a dielectric as a random medium described by a statistical ensemble of dipole configurations. As mentioned above, those configurations depend only on external d.o.f. and are independent of the atomic states. Now, on top of the quantum fluctuations that the dipoles and the EM field induce on each other, there are the stochastic fluctuations induced by the random distribution of scatterers. However, generally stochastic fluctuations do not enter the problem in the same fashion as quantum fluctuations. While the correlation time of the latter, $\tau$, satisfies $\tau\ll\omega_{AB}^{0}$, where $\hbar\omega_{AB}^{0}$ is the typical energy transferred in the internal processes, so that quantum fluctuations drive the dynamics of the atoms; the correlation time of the ensemble of configurations is much longer. Thus, the situation is generally that for quenched disorder and stochastic fluctuations enter only as performing ensemble averages --see also note 4. Let $P_{m}^{T}$ be the probability density of the mth dipole configuration at temperature $T$,
\begin{equation}
P_{m}^{T}\prod_{i=1}^{N}\textrm{d}^{3}R^{i}_{m}=\mathbf{Z}_{N}^{-1}(T)\prod_{i=1}^{N}\textrm{d}^{3}R^{i}_{m}\exp{[-U(\vec{R}_{m}^{1},..,\vec{R}_{m}^{N})/k_{B}T]},
\end{equation}
and $P_{m}^{T}|_{\vec{R}^{i}_{m}=\vec{r}}$ the conditional probability density with a dipole fixed at $\vec{r}$ --say the ith,
\begin{eqnarray}
P_{m}^{T}|_{\vec{R}^{i}_{m}=\vec{r}}\prod_{j=1}^{N}\textrm{d}^{3}R^{j}_{m}&=&\frac{\mathbf{Z}_{N}(T)}{\mathbf{Z}_{N-1}|_{\vec{R}^{i}=\vec{r}}(T)}\nonumber\\
&\times&\delta^{(3)}(\vec{r}-\vec{R}^{i}_{m})P_{m}^{T}\prod_{j=1}^{N}\textrm{d}^{3}R^{j}_{m},
\end{eqnarray}
with $\mathbf{Z}_{N-1}|_{\vec{R}^{i}=\vec{r}}(T)$ the conditional $(N-1)$-body partition function,
\begin{equation}
\mathbf{Z}_{N-1}|_{\vec{R}^{i}=\vec{r}}(T)=\int P_{m}^{T}\prod_{j=1}^{N}\textrm{d}^{3}R^{j}_{m}\delta^{(3)}(\vec{r}-\vec{R}^{i}_{m}).
\end{equation}
The statistical average of the expectation value of any local operator $\hat{\mathcal{O}}$ in the zero-temperature limit,
 $\mathcal{O}^{i}_{m}=\langle A^{i}_{m},\Omega_{m}|\hat{\mathcal{O}}|\Omega_{m},A^{i}_{m}\rangle$, reads,
\begin{equation}
\langle\mathcal{O}\rangle_{avg}^{T=0}(\vec{r})=\int P_{m}^{T=0}|_{\vec{R}^{i}_{m}=\vec{r}}\prod_{j=1}^{N}\textrm{d}^{3}R^{j}_{m}\mathcal{O}^{i}_{m}.
\end{equation}
The choice of $i$ is actually irrelevant since the dipoles are statistically equivalent and so must be any average observable. Alternatively, we can express symbolically $\langle\mathcal{O}\rangle_{avg}^{T=0}$ as the expectation value of $\hat{\mathcal{O}}$ in the stochastic polarized  EM vacuum and renormalized atomic ground state, $|\Omega_{avg},A_{avg}\rangle$.
Formally, this can be written in terms of the stationary reduced local (i.e. at each dipole location) density matrix of the system,
\begin{eqnarray}
\hat{\rho}^{T=0}(\vec{r})&=&\int P_{m}^{T=0}|_{\vec{R}^{i}_{m}=\vec{r}}\prod_{j=1}^{N}\textrm{d}^{3}R^{j}_{m}|\Omega_{m}\rangle\langle\Omega_{m}|\otimes
|A^{i}_{m}\rangle\langle A^{i}_{m}|\nonumber\\
&=&|\Omega_{avg}\rangle\langle\Omega_{avg}|\otimes|A_{avg}\rangle\langle A_{avg}|\Bigl|_{\vec{r}},
\end{eqnarray}
in which the integration amounts to taking the partial trace over the statistical mixture of dipole configuration states. The averaged expectation value of
$\hat{\mathcal{O}}$ at $T=0$ is
\begin{eqnarray}
\langle\mathcal{O}\rangle_{avg}^{T=0}(\vec{r})&=&
\int  P_{m}^{T=0}|_{\vec{R}^{i}=\vec{r}}\prod_{j=1}^{N}\textrm{d}^{3}R^{j}_{m}\langle\Omega^{m},A^{i}_{m}|
\hat{\mathcal{O}}|A^{i}_{m},\Omega^{m}\rangle\nonumber\\
&=&\langle\Omega_{avg},A_{avg}|\hat{\mathcal{O}}|A_{avg},\Omega_{avg}\rangle|_{\vec{r}}=\textrm{Tr}\{\hat{\rho}^{T=0}(\vec{r})
\cdot\hat{\mathcal{O}}\},\nonumber
\end{eqnarray}
where  $|A^{i}_{m},\Omega^{m}\rangle$ above depends implicitly on $\{\vec{R}^{j}_{m}\}$, $j=1,..,N$. If $\hat{\mathcal{O}}$ is a time-dependent interaction which couples weakly the dipoles to the EM field, by writing $\hat{\rho}$ as a direct product of EM and atomic states we assume that the EM field behaves as a stationary reservoir w.r.t. the weak interaction and the time-correlation between the reservoir and the dipoles is negligible as considering the dynamics of each dipole --cf. Ch.IV of \cite{Cohenbook}. Also, in a statistically homogeneous dielectric the spatial correlations functions are translation invariant and so are the stochastic states and average expectation values. Hence it suffices to compute them at a single point.

\subsection{Response functions and fluctuations}
Our goal is in the first place to give an expression for the Lamb shift of the atomic ground states, $\mathcal{E}^{LSh}$. Second, we will use that expression to compute the total binding energy of the ground state of the dielectric at zero temperature. Because the latter can be equally intended as a variation of the zero-point energy of the EM vacuum due to the presence of the dielectric --in the same spirit as the interpretation of Power \cite{Power}, Feynman \cite{Feynman} and Milonni \cite{MilonniScripta2} for the case of an only atom in free space-- we will refer to it as total vacuum energy density, $\mathcal{F}^{V}$. We will reinforce this interpretation giving an expression for $\mathcal{F}^{V}$ in terms of the source field propagator, in accordance to Schwinger's formalism.\\
\indent In order to compute the Lamb shifts and the total vacuum energy it is necessary to know the expressions for the two-time
quadratic correlation functions of the EM field and dipole operators in their renormalized vacuum and ground states respectively --cf. Ch.IV
of \cite{Cohenbook} and \cite{Dalibardetal}.
Generically, once the interaction Hamiltonian is turned on, ordinary second order time-dependent perturbation theory \cite{Sakurai} yields an energy shift in the ground state of each dipole, say the ith one with position vector $\vec{R}^{i}$, which corresponds to the generalized Lamb shift,\\
\begin{eqnarray}\label{laEShi}
\mathcal{E}^{LSh}_{\kappa,i}
&=&\hbar^{-1}\textrm{Tr}\Bigl\{\sum_{\gamma}|\langle \gamma,B^{i}_{\kappa}|e\hat{\vec{\mathbf{r}}}^{S}_{i}\cdot\hat{\vec{\mathbf{E}}}^{S}(\vec{R}^{i})|A^{i}_{\kappa},\Omega_{\kappa}\rangle|^{2}\Bigr\}\nonumber\\
&\times&\textrm{PV}\bigl[\frac{1}{\omega_{\gamma}+\omega^{\kappa,i}_{AB}}\bigr]\nonumber\\
&=&\hbar^{-1}\textrm{Tr}\Bigl\{\sum_{\gamma}|\langle\gamma|\hat{\vec{\mathbf{E}}}^{S}(\vec{R}^{i})|\Omega_{\kappa}\rangle|^{2}\cdot
|\langle A^{i}_{\kappa}|e\hat{\vec{\mathbf{r}}}^{S}_{i}|B^{i}_{\kappa}\rangle|^{2}\Bigr\}\nonumber\\
&\times&\textrm{PV}\bigl[\frac{1}{\omega_{\gamma}+\omega^{\kappa,i}_{AB}}\bigr].
\end{eqnarray}
In this formula $\{|\gamma\rangle\}$ is the set of intermediate excited EM states of energy $\hbar\omega_{\gamma}$,  $\omega^{\kappa,i}_{AB}$ is the transition frequency of the ith atom, the script $S$ stands for the time-independent operators in the Schr\"{o}dinger representation and the script $\kappa$ can take the values $0,m,avg$ corresponding to an only dipole in free space, to mth configuration of fixed dipoles and to a statistical ensemble of dipole configurations respectively. Using an appropriate identity for $\textrm{PV}\bigl[\frac{1}{\omega_{\gamma}+\omega^{\kappa,i}_{AB}}\Bigr]$, it is shown in Ch.IV
of \cite{Cohenbook} and \cite{Dalibardetal} that the above expression can be written in terms of quadratic correlators as,
\begin{eqnarray}
\mathcal{E}^{LSh}_{\kappa,i}&=&-(4\hbar)^{-1}\textrm{Tr}\Bigl\{\int_{-\infty}^{\infty}\textrm{d}\omega\Re{}\Bigl[
\langle\Omega_{\kappa}|\hat{\vec{\mathbf{E}}}(\vec{R}^{i};\omega)\otimes\hat{\vec{\mathbf{E}}}^{\dag}(\vec{R}^{i};\omega)|\Omega_{\kappa}\rangle\nonumber\\
&\cdot&\int\textrm{d}t\exp{[i\omega t]}i\Theta(t)\langle A^{i}_{\kappa}|[e\hat{\vec{\mathbf{r}}}_{i}(0),e\hat{\vec{\mathbf{r}}}_{i}(-t)]|A^{i}_{\kappa}\rangle\Bigr]\Bigr\}\label{vacfluc}\\
&-&(4\hbar)^{-1}\textrm{Tr}\Bigl\{\int_{-\infty}^{\infty}\textrm{d}\omega\Re{}\Bigl[
\langle A^{i}_{\kappa}|e\hat{\vec{\mathbf{r}}}_{i}(\omega)\otimes e\hat{\vec{\mathbf{r}}}^{\dag}_{i}(\omega)|A^{i}_{\kappa}\rangle\nonumber\\
&\cdot&
\int\textrm{d}t\exp{[i\omega t]}i\Theta(t)\langle\Omega_{\kappa}|[\hat{\vec{\mathbf{E}}}(\vec{R}^{i},0),\hat{\vec{\mathbf{E}}}(\vec{R}^{i},-t)]|\Omega_{\kappa}\rangle\Bigr]\Bigr\},
\label{backreact}
\end{eqnarray}
where Eq.(\ref{vacfluc}) is the energy shift associated to the polarization of the dipole due to the vacuum field fluctuations while Eq.(\ref{backreact}) is the energy shift due to the back-reaction of the polarized EM vacuum on the dipole. Therefore, the problem of computing $\mathcal{E}^{LSh}_{\kappa,i}$ reduces to calculate the equal point two-time quadratic correlation functions of the EM field and dipole operators in the corresponding EM vacuum and atomic ground state respectively \cite{Agarwal1}. In turn, the symmetric correlation function relates to the imaginary part of the linear response function via the fluctuation-dissipation theorem (FDT),
\begin{eqnarray}
\langle\Omega_{\kappa}|\hat{\vec{\mathbf{E}}}(\vec{R}^{i};\omega)&\otimes&\hat{\vec{\mathbf{E}}}^{\dag}(\vec{R}^{i};\omega)|\Omega_{\kappa}\rangle
=-\pi^{-1}\Im{}\Bigl\{\int\textrm{d}t\exp{[i\omega t]}i\Theta(t)\nonumber\\
&\times&\langle\Omega_{\kappa}|[\hat{\vec{\mathbf{E}}}(\vec{R}^{i},0),\hat{\vec{\mathbf{E}}}(\vec{R}^{i},-t)]|\Omega_{\kappa}\rangle\Bigr\},\label{FDTEM}\\
\langle A^{i}_{\kappa}|e\hat{\vec{\mathbf{r}}}_{i}(\omega)&\otimes& e\hat{\vec{\mathbf{r}}}^{\dag}_{i}(\omega)|A^{i}_{\kappa}\rangle=
-\pi^{-1}\Im{}\Bigl\{\int\textrm{d}t\exp{[i\omega t]}i\Theta(t)\nonumber\\
&\times&\langle A^{i}_{\kappa}|[e\hat{\vec{\mathbf{r}}}_{i}(0),e\hat{\vec{\mathbf{r}}}_{i}(-t)]|A^{i}_{\kappa}\rangle\Bigr\}\label{FDTalpha},
\end{eqnarray}
and the whole problem reduces to computing the local linear response functions. That is, the polarizability of the dipole in its ground state and the Green function of the electric field, which can be calculated classically. In the following we elaborate on the physical meaning of the response functions.
\subsubsection{An only dipole in free space}
\indent The equal point two-time commutator of the dipole operator for a two-level atom, say the ith one, in its bare ground state, $|A^{i}_{b}\rangle$, prior to the coupling to the radiation reservoir and at zero temperature reads,
\begin{eqnarray}\label{a}
&i&(\epsilon_{0}\hbar)^{-1}\Theta(t)
\langle A^{i}_{b}|[e\hat{\vec{\mathbf{r}}}_{i}(0),e\hat{\vec{\mathbf{r}}}_{i}(-t)]|A^{i}_{b}\rangle\nonumber\\
&=&i(\epsilon_{0}\hbar)^{-1}\Theta(t)\langle A_{b}^{i}|e\hat{\vec{\mathbf{r}}}_{i}^{S}|B^{i}_{b}\rangle\langle B^{i}_{b}|e\hat{\vec{\mathbf{r}}}_{i}^{S}
|A^{i}_{b}\rangle\nonumber\\
&\times&[\exp{(i\omega^{b,i}_{AB}t)}-\exp{(-i\omega^{b,i}_{AB}t)}].
\end{eqnarray}
The time-variable Fourier transform of the above equation is termed the bare dipole polarizability, $\bar{\alpha}_{b}(\omega)$ --for brevity we have omitted the location dependence,
\begin{eqnarray}\label{a}
\bar{\alpha}_{b}(\omega)&\equiv&2(\epsilon_{0}\hbar)^{-1}\langle A^{i}_{b}|e\hat{\vec{\mathbf{r}}}_{i}^{S}|B_{b}^{i}\rangle\langle B^{i}_{b}|e\hat{\vec{\mathbf{r}}}_{i}^{S}
|A^{i}_{b}\rangle\omega^{b,i}_{AB}[(\omega^{b,i}_{AB})^{2}-\omega^{2}]^{-1}\nonumber\\
&=&2(\epsilon_{0}\hbar)^{-1}\omega^{b,i}_{AB}[(\omega^{b,i}_{AB})^{2}-\omega^{2}]^{-1}
\vec{\mu}\otimes\vec{\mu},\textrm{ any $i$},
\end{eqnarray}
where  $\vec{\mu}=\langle A^{i}_{b}|e\hat{\vec{\mathbf{r}}}_{i}^{S}|B^{i}_{b}\rangle$ is the unique dipole-transition matrix element\footnote{It is implicit that in order to get 'bare' atomic bound states some EM modes have already been integrated out in the definition of $\omega^{b,i}_{AB}$.}. $\bar{\alpha}_{b}(\omega)$ is the response function of the dipole to a total monochromatic field acting upon $\vec{R}^{i}$ \footnote{The nomenclature used in this article varies slightly w.r.t. that in \cite{PRAvc}. In particular, $\alpha_{b}$ was denoted there as $\alpha'$. Also, what was referred to as self-polarization field in \cite{PRAvc} is referred in here as radiation-reaction field. In addition, instead of referring to its propagator as polarization propagator, we name it here EM field propagator in order to avoid confusion.}
\begin{equation}
\vec{p}^{\omega}(\vec{R}_{i})=\epsilon_{0}\bar{\alpha}_{b}(\omega)\cdot\vec{E}^{\omega}_{tot}(\vec{R}^{i}).
\end{equation}
For isotropic dipoles with a single oscillator --we can drop the dipole index,
\begin{eqnarray}
\bar{\alpha}_{b}(\omega)\equiv\alpha_{b}\bar{\mathbb{I}}
&=&\frac{2}{3}(\epsilon_{0}\hbar)^{-1}\omega^{b}_{AB}\mu^{2}[(\omega^{b}_{AB})^{2}-\omega^{2}]^{-1}\nonumber\\
&=&\frac{e^{2}}{3\epsilon_{0}m^{b}_{e}}[(\omega^{b}_{AB})^{2}-\omega^{2}]^{-1},\label{alpha0dee}
\end{eqnarray}
where $m^{b}_{e}$ is the bare electron mass.
However, since the effects of the EM vacuum fluctuations and radiation-reaction have not yet been incorporated, it does not satisfy the optical theorem.\\
\indent On the other hand, the time-variable Fourier transform of the electric field commutator in the EM vacuum of free space, $i\epsilon_{0}\hbar^{-1}\Theta(t)\langle\Omega_{0}|[\hat{\vec{\mathbf{E}}}(\vec{r},0),\hat{\vec{\mathbf{E}}}(\vec{r}',-t)]|\Omega_{0}\rangle$, is $(\omega/c)^{2}$ times the Green function of Maxwell's equation in free space, $\bar{G}^{(0)}(\vec{r}-\vec{r}';\omega)$,
\begin{equation}\label{Maxwellb}
\Bigl[\frac{\omega^{2}}{c^{2}}\bar{\mathbb{I}}-\vec{\nabla}\times\vec{\nabla}\times\Bigr]\bar{G}^{(0)}(\vec{r}-\vec{r}';\omega)
=\delta^{(3)}(\vec{r}-\vec{r}')\bar{\mathbb{I}}.
\end{equation}
$\bar{G}^{(0)}(\vec{r}-\vec{r}';\omega)$ is made of two contributions. These are, an electrostatic one,
\begin{equation}\label{stat}
\bar{G}_{stat.}^{(0)}(\vec{r};\omega)=\textrm{PV}\Bigl[\frac{1}{k^{2}}
\vec{\nabla}\otimes\vec{\nabla}\Bigr]\Bigl(\frac{-1}{4\pi\:r}\Bigr)+k^{-2}\bar{\mathbb{L}}\delta^{(3)}(\vec{r}),
\end{equation}
$k=\omega/c$, and a radiative field,
\begin{equation}\label{radi}
\bar{G}_{rad.}^{(0)}(\vec{r};\omega)=\frac{e^{i\:kr}}{-4\pi
r}\bar{\mathbb{I}}+\textrm{PV}\bigl[\frac{1}{k^{2}}\vec{\nabla}\otimes\vec{\nabla}\bigr]\frac{e^{i\:kr}-1}{-4\pi r},
\end{equation}
where the $\delta$-function in Eq.(\ref{stat}) must be intended in the sense of a distribution. The source tensor $\bar{\mathbb{L}}$ takes account of the geometry of the exclusion volume around each dipole-source. It satisfies Tr$\{\bar{\mathbb{L}}\}=1$ and, for a spherical  volume, $\bar{\mathbb{L}}=1/3\bar{\mathbb{I}}$ \cite{Japa}. $\bar{G}^{(0)}(\vec{r}-\vec{r}';\omega)$ is the response function of the EM field to a unique point dipole oscillating at frequency $\omega$ at $\vec{r}'$ in free space, $\vec{p}^{\omega}(\vec{r}')$, $\vec{E}^{\omega}(\vec{r})=k^{2}\epsilon_{0}^{-1}\bar{G}^{(0)}(\vec{r}-\vec{r}';\omega)\cdot\vec{p}^{\omega}(\vec{r}')$. In momentum space the radiative component is totally transverse w.r.t. the propagation direction while the electrostatic one is fully longitudinal,
\begin{equation}
\bar{G}^{(0)}(\vec{q};\omega)=G_{\perp}^{(0)}(q)\bar{\textrm{P}}_{\perp}(\hat{q})+G_{\parallel}^{(0)}(q)\bar{\textrm{P}}_{\parallel}(\hat{q}),
\end{equation}
with
\begin{equation}\label{losfreek}
G_{\perp}^{(0)}(q)=\frac{1}{k^{2}-q^{2}},\quad G_{\parallel}^{(0)}(q)=\frac{1}{k^{2}}.
\end{equation}
and $\bar{\textrm{P}}_{\perp}(\hat{q})=\bar{\mathbb{I}}-\hat{q}\otimes\hat{q}$,  $\bar{\textrm{P}}_{\parallel}(\hat{q})=\hat{q}\otimes\hat{q}$ being the transverse and longitudinal projectors respectively, with $\hat{q}$ the unitary vector in the direction of propagation. Hereafter and for the sake of brevity we will drop the explicit $\omega$ and/or $q$ dependence from the functional arguments unless necessary.\\
\indent When the ith dipole, isolated, couples to radiation via the interaction Hamiltonian, the aforementioned radiative reaction of the EM field in free space renormalizes the atomic states to $|A_{0}^{i}\rangle,|B_{0}^{i}\rangle$ and, correspondingly, the single particle polarizability. Following a simple renormalization procedure --cf.\cite{PRAvc,deVriesPRL,Felderhof3} we obtain,
\begin{equation}\label{a}
\bar{\alpha}(\omega)\equiv\bar{\alpha}_{0}\Bigl(1+ik^{2}\textrm{Tr}\{\bar{\alpha}_{0}\cdot\Im{[\bar{G}^{(0)}(\vec{r},\vec{r};\omega)]}\}
\Bigr)^{-1},
\end{equation}
where the divergent real part of $\bar{G}^{(0)}(\vec{r},\vec{r};\omega)$ has been incorporated in the renormalization of the free electron mass and the free space resonant frequency, $m_{e},\omega^{0}_{AB}$, so that $\alpha_{0}\equiv\frac{e^{2}}{3\epsilon_{0}m_{e}}[(\omega^{0}_{AB})^{2}-\omega^{2}]^{-1}$. Hereafter we will denote $\omega^{0}_{AB}$ simply by $\omega_{0}$.
$\bar{\alpha}(\omega)$ is the response function of a unique dipole in free space to an incident monochromatic probe field acting, say, upon $\vec{R}^{i}$,
\begin{equation}
\vec{p}^{\omega}(\vec{R}^{i})=\epsilon_{0}\bar{\alpha}(\omega)\cdot\vec{E}^{\omega}_{inc}(\vec{R}^{i}).
\end{equation}
\subsubsection{A specific dipole configuration}
Next, when all the dipoles are brought together, it was shown in \cite{PRAvc} how a classical diagrammatic renormalization procedure leads to renormalized values for the single particle polarizability and the EM field propagator. In the first place, let us consider a fixed configuration of dipoles, $m$. The polarization propagator reads,
\begin{equation}
\bar{\Pi}_{m}^{\omega}(\vec{r},\vec{r}^{'})=\sum_{i,j=0}^{N}\bar{\mathrm{\pi}}_{m}^{\omega}(\vec{R}^{i}_{m},\vec{R}^{j}_{m})\:\delta^{(3)}(\vec{r}-\vec{R}^{i}_{m})\delta^{(3)}(\vec{r}^{'}-\vec{R}^{j}_{m}),
\end{equation}
where $\bar{\mathrm{\pi}}_{m}^{\omega}(\vec{R}^{i}_{m},\vec{R}^{j}_{m})$ is
\begin{eqnarray}\label{polaris}
\bar{\mathrm{\pi}}_{m}^{\omega}(\vec{R}^{i}_{m},\vec{R}^{j}_{m})&=&\int \textrm{d}t e^{i\omega t}i(\epsilon_{0}\hbar)^{-1}\Theta(t)
\langle A_{m}^{i}|[e\hat{\vec{\mathbf{r}}}_{i}(0),e\hat{\vec{\mathbf{r}}}_{j}(-t)]|A_{m}^{j}\rangle\nonumber\\
&=&[\alpha_{0}^{-1}\bar{\mathbb{I}}\delta_{ij}+k^{2}\bar{G}^{(0)'}(\vec{R}^{i}_{m},\vec{R}^{j}_{m})]^{-1}.
\end{eqnarray}
$\bar{G}^{(0)'}(\vec{R}^{i}_{m},\vec{R}^{j}_{m})\equiv\bar{G}^{(0)}(\vec{R}^{i}_{m},\vec{R}^{j}_{m})
-\Re{\{\bar{G}^{(0)}(\vec{R}^{i}_{m},\vec{R}^{j}_{m})\}}\delta_{ij}$ is defined to take account of the regularization of the real divergence in $\bar{G}^{(0)}(\vec{R}^{i}_{m},\vec{R}^{i}_{m})$. The inversion in Eq.(\ref{polaris}) must be intended w.r.t. the particle indices, $i,j$. In particular, the local term proportional to $\delta_{ij}$ is defined as the ith renormalized polarizability, $\bar{\breve{\alpha}}_{m}^{i}(\omega)$. It is made of an infinite series of multiple-scattering diagrams which start and end at the ith dipole. $\bar{\mathrm{\pi}}_{m}^{\omega}(\vec{R}^{i}_{m},\vec{R}^{j}_{m})$ is the (non-local) response function of the dielectric to a generic monochromatic external electric field,
\begin{equation}
\vec{p}^{\omega}(\vec{R}^{i}_{m})=\epsilon_{0}\sum_{j}^{N}\bar{\mathrm{\pi}}_{m}^{\omega}(\vec{R}^{i}_{m},\vec{R}^{j}_{m})
\cdot\vec{E}^{\omega}_{ext}(\vec{R}^{j}_{m}).
\end{equation}
The relation of $\bar{\mathrm{\pi}}_{m}^{\omega}$ with the  $\mathfrak{t}$-matrix defined in the so-called Coupled Dipole Method (CDM) \cite{Sentenac} is $\bar{\mathrm{\pi}}_{m}^{\omega}=-k^{-2}\bar{\mathfrak{t}}^{\omega}_{m}$.\\
\indent Next, let us compute the EM field propagator which yields the radiative corrections on $\bar{\breve{\alpha}}_{m}^{i}(\omega)$, $\bar{\mathfrak{g}}_{m}(\vec{r},\vec{R}_{m}^{i};\omega)$. It is given by an equation similar to that of Maxwell's in free space,
\begin{equation}\label{Polstoch}
\Bigl[\frac{\omega^{2}}{c^{2}}e^{\omega}_{m,i}(\vec{r})\bar{\mathbb{I}}-\vec{\nabla}\times\vec{\nabla}\times\Bigr]\bar{\mathfrak{g}}_{m}(\vec{r},\vec{R}^{i}_{m};\omega)
=\delta^{(3)}(\vec{r}-\vec{R}^{i}_{m})\bar{\mathbb{I}},
\end{equation}
with $\tilde{e}^{\omega}_{m}(\vec{r})=1+\alpha_{0}\sum_{j=1,N}\delta^{(3)}(\vec{r}-\vec{R}^{j}_{m})$ and
$e^{\omega}_{m,i}(\vec{r})= \tilde{e}^{\omega}_{m}(\vec{r})-\alpha_{0}\delta^{(3)}(\vec{r}-\vec{R}^{i}_{m})$.
The source fixed at the position vector of the ith scatterer is removed from the permittivity function on the l.h.s. of the equation. The radiation-reaction propagator is computed out of $\bar{\mathfrak{g}}_{m}$ making the source and the emitter coincide. It was calculated in \cite{PRAvc} in function of the polarization propagator,
\begin{equation}\label{Gdiscr}
\bar{\mathfrak{g}}_{m}(\vec{R}^{i}_{m},\vec{R}^{i}_{m};\omega)=\sum_{j=0}^{N}
\bar{G}^{(0)}(\vec{R}_{m}^{i}-\vec{R}_{m}^{j})\cdot\bar{\mathrm{\pi}}_{m}^{\omega}(\vec{R}^{j}_{m},\vec{R}^{i}_{m})[\bar{\breve{\alpha}}_{m}^{i}]^{-1}.
\end{equation}
and by consistency,
\begin{equation}
\bar{\breve{\alpha}}_{m}^{i}=\bar{\alpha}_{0}[1+k^{2}
\textrm{Tr}\{\bar{\mathfrak{g}}^{'}_{m}(\vec{R}^{i}_{m},\vec{R}^{i}_{m};\omega)\cdot\bar{\alpha}_{0}\}]^{-1},
\end{equation}
where again $\bar{\mathfrak{g}}^{'}_{m}(\vec{R}^{i}_{m},\vec{R}^{i}_{m};\omega)\equiv
\bar{\mathfrak{g}}_{m}(\vec{R}^{i}_{m},\vec{R}^{i}_{m};\omega)-\Re{\{\bar{G}^{(0)}(\vec{R}^{i}_{m},\vec{R}^{i}_{m};\omega)\}}$.\\
\indent Applying reciprocity, $\bar{\mathfrak{g}}_{m}(\vec{R}^{i}_{m},\vec{r}';\omega)$ yields the incident field which reaches a host dipole at $\vec{R}^{i}_{m}$, whose source is a non-polarizable external monochromatic dipole at any point $\vec{r}'$, $\vec{\mu}_{ext}^{\omega}$,
\begin{equation}
\vec{E}^{\omega}_{inc}(\vec{R}_{m}^{i})=k^{2}\epsilon_{0}^{-1}\bar{\mathfrak{g}}_{m}(\vec{R}^{i}_{m},\vec{r}';\omega)\cdot\vec{\mu}_{ext}^{\omega}.
\end{equation}
Note that since the host dipole at $\vec{R}_{m}^{i}$ is polarizable, the total field at $\vec{R}_{m}^{i}$ contains an additional contribution from the radiation-reaction (rr) field. As a result we have,
\begin{eqnarray}
\vec{E}^{\omega}_{tot}(\vec{R}_{m}^{i})&\equiv&\vec{E}^{\omega}_{inc}(\vec{R}_{m}^{i})+\vec{E}^{\omega}_{rr}(\vec{R}_{m}^{i})\nonumber\\
&=&k^{2}\epsilon_{0}^{-1}\bar{\alpha}^{-1}_{0}\cdot\bar{\breve{\alpha}}_{m}^{i}\cdot
\bar{\mathfrak{g}}_{m}(\vec{R}^{i}_{m},\vec{r}';\omega)\cdot\vec{\mu}^{\omega}_{ext}.
\end{eqnarray}
From the above equation we can identify a new propagator which accounts for both the two-point, two-time commutator of the EM field in vacuum and the radiation reaction field fluctuations which dress up the renormalized polarizability of the emitter/source dipole,
\begin{equation}\label{gtilde}
\bar{\tilde{\mathfrak{g}}}_{m}(\vec{R}^{i}_{m},\vec{r}';\omega)\equiv\Bigl[\bar{\mathbb{I}}+k^{2}
\bar{\mathfrak{g}}^{'}_{m}(\vec{R}^{i}_{m},\vec{R}^{i}_{m})\cdot\bar{\alpha}_{0}\Bigr]^{-1}\cdot\bar{\mathfrak{g}}_{m}(\vec{R}^{i}_{m},\vec{r}'),
\end{equation}
such that
\begin{equation}
\vec{E}^{\omega}_{tot}(\vec{R}_{m}^{i})=k^{2}\epsilon_{0}^{-1}\bar{\tilde{\mathfrak{g}}}_{m}(\vec{R}^{i}_{m},\vec{r}';\omega)\cdot\vec{\mu}^{\omega}_{ext}.
\end{equation}
$\bar{\tilde{\mathfrak{g}}}_{m}$ is the Green function of an equation like that of Eq.(\ref{Polstoch}) but replacing $e^{\omega}_{m,i}(\vec{r})$ there with the total dielectric function $\tilde{e}^{\omega}_{m}(\vec{r})$.
\subsubsection{Random medium}
\begin{figure}[h]
\includegraphics[height=8.4cm,width=8.2cm,clip]{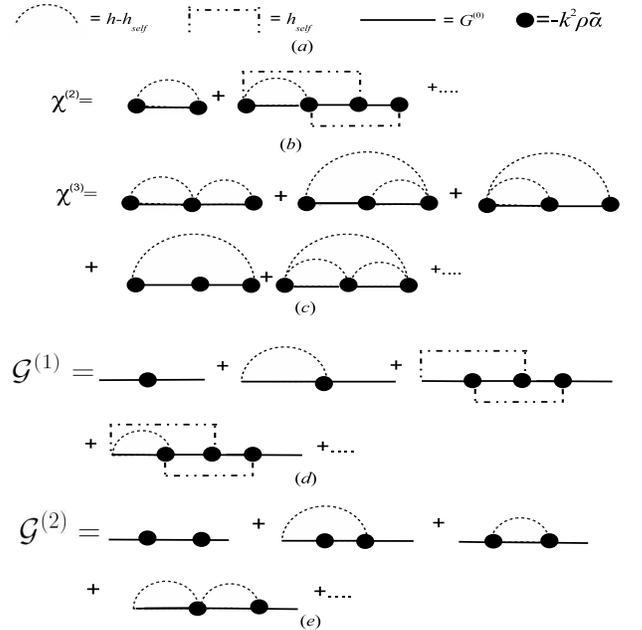}
\caption{($a$) Diagrammatic representation of Feynman's rules. Only two-point correlation functions are considered for simplicity. The self-correlation function, $h_{self}(\vec{r})=\rho^{-1}\delta^{(3)}(\vec{r})$, appears separated from the rest. $(b),(c)$ Examples of 1PI diagrams which amount to $\chi^{(2)}$ and $\chi^{(3)}$ respectively. The lowest order recurrent scattering diagram, $\chi^{(2,2)}$, is included in the series of $\chi^{(2)}$. $(d),(e)$ Examples of order $\rho$ and order $\rho^{2}$ multiple-scattering diagrams which amount to $\mathcal{G}^{(1)}$ and $\mathcal{G}^{(2)}$ respectively. The third and fourth diagrams in $(d)$, which amount to $\mathcal{G}^{(1,2)}$. are entangled recurrent scattering diagrams. In them, the self-correlation functions which affect the first and second scatterers cannot be factored out.}\label{fig1}
\end{figure}

Finally, let us consider a statistical ensemble of configurations of statistically equivalent isotropic dipoles. By simply taking the statistical average on the precedent equations,
\begin{equation}
\bar{\mathcal{G}}(\vec{r},\vec{r}';\omega)=\int P_{m}^{T=0}|_{\vec{R}^{i}_{m}=\vec{r}}\prod_{j=1}^{N}\textrm{d}^{3}R^{j}_{m}
\bar{\mathfrak{g}}_{m}(\vec{R}^{i}_{m},\vec{r}';\omega),
\end{equation}
 or using diagrammatical techniques otherwise in momentum space --cf. \cite{PRAvc}, we end up with
\begin{eqnarray}
&i&\frac{\epsilon_{0}}{k^{2}\hbar}\int \textrm{d}t\textrm{d}^{3}r\exp{[i(\omega t+\vec{q}\cdot\vec{r})]}\Theta(t)\nonumber\\
&\times&\langle\Omega_{avg}|[\hat{\vec{\mathbf{E}}}(\vec{0},0),\hat{\vec{\mathbf{E}}}(\vec{r},-t)]|\Omega_{avg}\rangle\nonumber\\
&=&\mathcal{G}_{\perp}(q;\omega)(\bar{\mathbb{I}}-\hat{q}\otimes\hat{q})+
\mathcal{G}_{\parallel}(q;\omega)\:\hat{q}\otimes\hat{q},\nonumber
\end{eqnarray}
where
\begin{eqnarray}
\mathcal{G}_{\perp}(q;\omega)&=&\frac{\chi_{\perp}(q;\omega)}{\rho\tilde{\alpha}}G_{\perp}(q;\omega)=
\frac{\chi_{\perp}(q;\omega)/(\rho\tilde{\alpha})}{k^{2}[1+\chi_{\perp}(q;\omega)]-q^{2}},\nonumber\\
\mathcal{G}_{\parallel}(q;\omega)&=&\frac{\chi_{\parallel}(q;\omega)}{\rho\tilde{\alpha}}\:G_{\parallel}(q;\omega)
=\frac{1}{\rho\tilde{\alpha}}\:\frac{\chi_{\parallel}(q;\omega)}{k^{2}[1+\chi_{\parallel}(q;\omega)]}.\label{LDOSIparal}
\end{eqnarray}
and $G_{\perp,\parallel}(q;\omega)$ are the transverse and longitudinal components of the Dyson (bulk) propagators readily identifiable from the second equalities. The stochastic renormalized polarizability, $\tilde{\alpha}$, is defined later on. $\chi_{\perp,\parallel}(q;\omega)$ are the transverse and longitudinal components of the electrical susceptibility. Diagrammatically, it is made of a series of one-particle-irreducible (1PI) scattering processes like those in Figs.\ref{fig1}$(b,c)$. The relationship of proportionality between the polarization and Dyson's propagator in $q$-space allows us to define the \emph{local field factors},  $\mathcal{L}_{\perp,\parallel}(q)=\frac{\chi_{\perp,\parallel}(q)}{\rho\tilde{\alpha}}$, such that $\mathcal{G}_{\perp,\parallel}=\mathcal{L}_{\perp,\parallel}G_{\perp,\parallel}$. Alternatively, Eq.(\ref{LDOSIparal}) can be written also as,
\begin{equation}\label{infunctionofG}
\mathcal{G}_{\perp,\parallel}(q)=\frac{1}{k^{2}\rho\tilde{\alpha}}
\Bigl[1-\frac{G_{\perp,\parallel}}{G_{\perp,\parallel}^{(0)}}\Bigr].
\end{equation}
Equivalently, in terms of the stochastic polarization propagator,
\begin{eqnarray}
\bar{\Pi}^{\omega}(\vec{r},\vec{r}^{'})&=&\int \frac{\mathbf{Z}_{N}(0)}{\mathbf{Z}_{N-2}|^{\vec{R}^{j}=\vec{r}'}_{\vec{R}^{i}=\vec{r}}(0)} P_{m}^{T=0}\prod_{l=1}^{N}\textrm{d}^{3}R^{l}_{m}\bar{\mathrm{\pi}}_{m}^{\omega}(\vec{R}^{i}_{m},\vec{R}^{j}_{m})\nonumber\\
&\times&\delta^{(3)}(\vec{r}-\vec{R}^{i}_{m})\delta^{(3)}(\vec{r}^{'}-\vec{R}^{j}_{m}),
\end{eqnarray}
\begin{equation}
\textrm{with }\mathbf{Z}_{N-2}|^{\vec{R}^{j}=\vec{r}'}_{\vec{R}^{i}=\vec{r}}(T)=\int P_{m}^{T}\prod_{l=1}^{N}\textrm{d}^{3}R^{l}_{m}\delta^{(3)}(\vec{r}-\vec{R}^{i}_{m})
\delta^{(3)}(\vec{r}'-\vec{R}^{j}_{m}),\nonumber
\end{equation}
we may write
\begin{equation}\label{latercera}
\mathcal{G}_{\perp,\parallel}(q;\omega)=(\rho\tilde{\alpha})^{-1}G_{\perp,\parallel}^{(0)}(q;\omega)\Pi^{\omega}_{\perp,\parallel}(q).
\end{equation}
For convenience we define a scalar radiative potential in terms of the trace of the radiation-reaction propagator, $\phi(\omega)=\phi^{(0)}(\omega)+\phi^{sc}(\omega)$, where,
\begin{eqnarray}
\phi^{(0)}(\omega)&\equiv&\frac{\omega^{2}}{c^{2}}i\Im{\{2\int\frac{\textrm{d}^{3}q}{(2\pi)^{3}}G_{\perp}^{(0)}\}}=\frac{-\omega^{3}}{2\pi c^{3}}i,\label{42a}\\
\phi^{sc}(\omega)&\equiv&\frac{\omega^{2}}{c^{2}}\int\frac{\textrm{d}^{3}q}{(2\pi)^{3}}\Bigl[2\Bigl(\mathcal{G}_{\perp}-G_{\perp}^{(0)}\Bigr)
\nonumber\\&+&\Bigl(\mathcal{G}_{\parallel}-G_{\parallel}^{(0)}\Bigr)\Bigr]\equiv 2\phi^{sc}_{\perp}+\phi^{sc}_{\parallel}.
\end{eqnarray}
For further convenience, we have distinguished the contribution of transverse from that of longitudinal modes in $\phi^{sc}$.
The script $sc$ stands for \emph{scattering} since $\phi^{sc}(\omega)$ is the part of $\phi$ which involves multiple scattering processes. In function of $\phi$, $\alpha$ and the renormalized stochastic polarizability appearing in Eqs.(\ref{LDOSIparal},\ref{latercera}) read,
\begin{eqnarray}
\alpha&=&\alpha_{0}(1+\alpha_{0}\phi^{(0)})^{-1},\label{alphaU0}\\
\tilde{\alpha}&=&\alpha_{0}(1+\alpha_{0}\phi)^{-1}=\alpha(1+\alpha \phi^{sc})^{-1},\label{alphaU}
\end{eqnarray}
they all being isotropic, $\bar{\alpha}_{0}=\alpha_{0}\bar{\mathbb{I}}$, $\bar{\alpha}=\alpha\bar{\mathbb{I}}$, $\bar{\tilde{\alpha}}=\tilde{\alpha}\bar{\mathbb{I}}$. When a single dipole is excited by a monochromatic external field, the emitted power is given by $W^{\omega}=\frac{\omega\epsilon_{0}}{2}|\vec{E}^{\omega}_{ext}|^{2}\Im{\{\tilde{\alpha}\}}$, in agreement with the optical theorem. By parameterizing $\alpha$ and $\tilde{\alpha}$ as Lorentzian polarizabilities, formulas were found in \cite{PRAvc} for the decay rate, $\Gamma$, and frequency shifts. In particular, in free space, $\Gamma_{0}=\frac{k_{0}^{3}\mu^{2}}{3\pi\epsilon_{0}\hbar}$.\\
\indent As for the case of a specific dipole configuration it is possible to define a propagator which accounts for both the two-point, two-time commutator of the EM field in the stochastic vacuum and the radiation reaction field fluctuations which dress up the renormalized stochastic polarizability,
\begin{equation}
\bar{\tilde{\mathcal{G}}}(\vec{r},\vec{r}';\omega)\equiv\bar{\mathcal{G}}(\vec{r},\vec{r}';\omega)(1+\alpha_{0}\phi)^{-1}.\label{laGstochSch}
\end{equation}
The total averaged electric field which acts upon a host dipole at $\vec{r}$ and whose source is an external monochromatic dipole sited at $\vec{r}'$ reads,
\begin{equation}
\langle\vec{E}^{\omega}_{tot}(\vec{r})\rangle_{avg}=k^{2}\epsilon_{0}^{-1}\bar{\tilde{\mathcal{G}}}(\vec{r},\vec{r}';\omega)\cdot\vec{\mu}^{\omega}_{ext}.
\end{equation}
\indent For the specific computations in a random medium we will use a renormalization scheme similar to that of Felderhof and Cichocki \cite{Felderhof3}. It consists of two complementary steps which reflect the mutual polarization of the dipoles and the EM reservoir. In the first one, the polarizabilities are renormalized by an infinite number of radiation-reaction cycles as outlined above. Diagrammatically, those processes amount to recurrent scattering terms in which the initial and final scatterers coincide. Note that those diagrams --cf. Fig.\ref{fig1}$(d)$-- may contain entangled intermediate recurrent scattering processes signaled by self-correlation functions. In the second step, the rest of 1PI diagrams which are not accounted for in the renormalization of the polarizability are added up in the electrical susceptibility. Again, these diagrams may contain also entangled intermediate recurrent scattering processes --cf. Figs.\ref{fig1}$(b,c)$. Consistency between both steps is guaranteed by demanding that the scatterers in all those diagrams be renormalized. Bearing in mind that unentangled recurrent scattering process are all accounted for in the definition of $\tilde{\alpha}$, the electrical susceptibility components,  $\chi_{\perp,\parallel}(q;\omega)$, adjust to cluster expansions of the form,
\begin{equation}\label{laXenAes}
\chi_{\perp,\parallel}(q;\omega)=\sum_{n=1}\chi_{\perp,\parallel}^{(n)}(q;\omega)=\sum_{n=1,m=0}^{\infty}X^{(n,m)}_{\perp,\parallel}(q;\omega)\rho^{n}
\tilde{\alpha}^{n+m}.
\end{equation}
The functions
$X^{(n,m)}_{\perp,\parallel}(q;\omega)$  incorporate the spatial dispersion due to the 1PI spatial correlations within clusters of $n$ scatterers in which all the self-correlation functions appear entangled. The index $m$ stands for the total multiplicity of entangled intermediate recurrent scattering events. The same kind of decomposition is applicable to $\mathcal{G}_{\perp,\parallel}$ and $\phi^{sc}$ --cf. Fig.\ref{fig1}.\\

\section{The Lamb shift and the total vacuum energy}\label{S3}
\subsection{Lamb shift and Lamb energy in free space}
We are now in conditions to compute all the physical quantities. We start with the Lamb shift in free space. Applying the FDT on  Eqs.(\ref{vacfluc},\ref{backreact}) with $\kappa=0$ and substituting there Eqs.(\ref{42a},\ref{alphaU0}) we obtain the free-space Lamb-shift\footnote{We emphasize that in the computation of any local physical observable which is a combination of dipole and EM field operators the inherent divergence in $\Re{\{\bar{G}^{(0)}(\vec{R}^{i}_{m},\vec{R}^{i}_{m};\omega)\}}$ must be intended as regularized in both the free electron mass and the free-space resonant frequency of the oscillator. Hence, the use of  regularized "primed" field propagators.} \cite{Wylie},
\begin{equation}
\mathcal{E}^{LSh}_{0}=\frac{\hbar}{2\pi}\int_{0}^{\infty}\textrm{d}\omega\:\Im{}\{\alpha_{0}(1+\alpha_{0}\phi^{(0)})^{-1}\phi^{(0)}\}.\label{eer23}
\end{equation}
From Eq.(\ref{eer23}) we define the density of states contributing to the free-space shift, $\mathcal{N}^{LSh}_{0}(\omega)=\Im{}\{\frac{\rho}{2\pi}\alpha_{0}\phi^{(0)}(1+\alpha_{0}\phi^{(0)})^{-1}\}$.\\
\indent Next, using the Feynman-Pauli theorem --cf. pp.295-297 of \cite{PinesNoezius}, the free-space vacuum energy is the energy gained by the system atoms-reservoir as the interaction Hamiltonian is turned on adiabatically neglecting dipole mutual couplings. This is parameterized by varying the coupling constant squared from zero to its actual value $e^{2}$. Since $\alpha_{0}$ is quadratic in $e$, we can write,
\begin{eqnarray}
\mathcal{F}^{V}_{0}&=&\int_{0}^{\infty}\hbar\textrm{d}\omega\int_{0}^{\alpha_{0}}\frac{\delta\alpha^{'}_{0}}{\alpha^{'}_{0}}\mathcal{N}^{LSh}_{0}=
\frac{\rho\hbar}{2\pi}\int_{0}^{\infty}\textrm{d}\omega\Im{\{\ln{[1+\alpha_{0}\phi]}\}}\nonumber\\
&=&-\frac{\rho\hbar}{2\pi}\int_{0}^{\infty}\textrm{d}\omega\:\Im{\{\ln{[\alpha/\alpha_{0}]}\}},\label{eer223}
\end{eqnarray}
result that was firstly obtained by Agarwal \cite{Agarwal2}. The free-space energies are additive and so we will refer to $\mathcal{F}^{V}_{0}$  also as free-space Lamb energy density, $\mathcal{F}^{L}_{0}$.

\subsection{Lamb shift and total vacuum energy of a fixed configuration}
Let us consider the mth configuration of dipoles. Following analogous steps to those for $\mathcal{E}^{LSh}_{0}$ and $\mathcal{F}^{V}_{0}$, we obtain instead,
\begin{eqnarray}
\mathcal{E}^{LSh,i}_{m}&=&\frac{\hbar}{2\pi}\int_{0}^{\infty}\textrm{d}\omega\:\Im{}\{
\textrm{Tr}[\bar{\breve{\alpha}}_{m}^{i}\cdot\bar{\mathfrak{g}}^{'}_{m}(\vec{R}^{i}_{m},\vec{R}^{i}_{m})]\}\label{eer24}\\
&=&\frac{\hbar}{2\pi}\int_{0}^{\infty}\textrm{d}\omega\sum_{j=1}^{N}
\Im{}\{\textrm{Tr}[\bar{G}^{(0)'}(\vec{R}_{m}^{i},\vec{R}_{m}^{j})\cdot\bar{\mathrm{\pi}}_{m}^{\omega}(\vec{R}^{j}_{m},\vec{R}^{i}_{m})]\},\nonumber
\end{eqnarray}
where the trace operation applies over spatial tensor indices only. The first expression after the equality symbol was obtained by Buhmann \emph{et al.} \cite{WelschI,WelschII} using a fully quantum-mechanical formalism. Making the identification
\begin{equation}
\mathcal{N}^{LSh,i}_{m}=\mathcal{V}^{-1}\sum_{j=1}^{N}
\Im{}\{\textrm{Tr}[\bar{G}^{(0)'}(\vec{R}_{m}^{i},\vec{R}_{m}^{j})\cdot\bar{\mathrm{\pi}}_{m}^{\omega}(\vec{R}^{j}_{m},\vec{R}^{i}_{m})]\},
\end{equation}
with the trace again intended over spatial indices only, the total vacuum energy density is
\begin{eqnarray}
\mathcal{F}^{V}_{m}&=&\sum_{i=1}^{N}\int_{0}^{\infty}\hbar\textrm{d}\omega\int_{0}^{\alpha_{0}}\frac{\delta\alpha^{'}_{0}}{\alpha^{'}_{0}}
\mathcal{N}^{LSh,i}_{m}\nonumber\\
&=&\frac{\hbar}{2\pi\mathcal{V}}\int_{0}^{\infty}\textrm{d}\omega
\Im{}\Bigl\{\textrm{Tr}\Bigl[\ln{\Bigl(\bar{\mathbb{I}}\delta_{ij}+k^{2}\bar{\alpha}_{0}
\cdot\bar{G}^{(0)'}(\vec{R}^{i}_{m},\vec{R}^{j}_{m})\Bigr)\Bigr]}\Bigr\},\label{eer224}
\end{eqnarray}
where the trace operation applies both over dipole indices $(i,j)$ and spatial tensor components.
This result was firstly obtained by Renne \cite{Rene} making the sum over normal modes and then by Agarwal \cite{Agarwal2} using the FDT in an approach very similar to ours, and more recently by Emig \emph{et al.} \cite{JaffePRD} using a variant of the Schwinger's source theory. Its expansion in multiple-scattering interactions is,
\begin{eqnarray}
\mathcal{F}^{V}_{m}&=&\mathcal{F}^{L}_{0}+\frac{\hbar}{2\pi\mathcal{V}}\int_{0}^{\infty}\textrm{d}\omega
\Im{}\Bigl\{\sum_{n=1}^{\infty}\frac{(-1)^{n+1}}{n}\textrm{Tr}
\{[k^{2}\bar{\alpha}\nonumber\\&\cdot&\bar{G}^{(0)''}(\vec{R}^{i}_{m},\vec{R}^{j}_{m})]^{n}\}\Bigr\}\label{manybody}\\
&=&\mathcal{F}^{L}_{0}+\frac{\hbar}{2\pi\mathcal{V}}\int_{0}^{\infty}\textrm{d}\omega
\Im{}\Bigl\{\frac{-1}{2}\textrm{Tr}\{(k^{2}\bar{\alpha})^{2}\sum_{i\neq j}\bar{G}^{(0)}(\vec{R}^{j}_{m},\vec{R}^{i}_{m})\nonumber\\&\cdot&
\bar{G}^{(0)}(\vec{R}^{i}_{m},\vec{R}^{j}_{m})\}+\frac{1}{3}\textrm{Tr}\{(k^{2}\bar{\alpha})^{3}\sum_{i\neq j\neq l\neq i}\bar{G}^{(0)}(\vec{R}^{i}_{m},\vec{R}^{j}_{m})\nonumber\\&\cdot&
\bar{G}^{(0)}(\vec{R}^{j}_{m},\vec{R}^{l}_{m})\cdot
\bar{G}^{(0)}(\vec{R}^{l}_{m},\vec{R}^{i}_{m})\}+..\Bigr\},\nonumber\\
&\textrm{where }&\bar{G}^{(0)''}(\vec{R}^{i}_{m},\vec{R}^{j}_{m})=\bar{G}^{(0)}(\vec{R}^{i}_{m},\vec{R}^{j}_{m})
-\bar{G}^{(0)}(\vec{R}^{i}_{m},\vec{R}^{j}_{m})\delta_{ij}\nonumber
\end{eqnarray}
\indent In order to get a deeper insight into the physical interpretation of the above formula, we can write it in three equivalent manners,
\begin{eqnarray}
\mathcal{F}^{V}_{m}&=&
-\frac{\hbar}{2\pi\mathcal{V}}\int_{0}^{\infty}\textrm{d}\omega
\Im{}\Bigl\{\textrm{Tr}\Bigl[\ln{[\bar{\mathrm{\pi}}_{m}^{\omega}(\vec{R}^{j}_{m},\vec{R}^{i}_{m})/\alpha_{0}]}\Bigr]\Bigr\}\label{la1}\\
&=&-\frac{\hbar}{2\pi\mathcal{V}}\int_{0}^{\infty}\textrm{d}\omega
\Im{}\Bigl\{\textrm{Tr}\Bigl[\ln{}[\bar{\tilde{\mathfrak{g}}}_{m}(\vec{R}^{i}_{m},\vec{R}^{j}_{m})
\nonumber\\&\cdot&[\bar{G}^{(0)}]^{-1}(\vec{R}^{i}_{m},\vec{R}^{j}_{m})]\Bigr]\Bigr\}\label{la2}\\
&=&
-\frac{\hbar}{2\pi\mathcal{V}}\int_{0}^{\infty}\textrm{d}\omega
\Im{}\Bigl\{\textrm{Tr}\Bigl[\ln{}[\bar{\breve{\alpha}}_{m}^{i}\cdot\bar{\mathfrak{g}}_{m}(\vec{R}^{i}_{m},\vec{R}^{j}_{m})
\nonumber\\&\cdot&\bar{\alpha}_{0}^{-1}\cdot[\bar{G}^{(0)}]^{-1}(\vec{R}^{i}_{m},\vec{R}^{j}_{m})]\Bigr]\Bigr\}.\label{la3}
\end{eqnarray}
which shows that  $\mathcal{F}^{V}_{m}$ can be expressed in function of atomic d.o.f. only [Eq.(\ref{la1})], EM d.o.f. only [Eq.(\ref{la2})] or as a combination of both [Eq.(\ref{la3})]. In particular, the expression in Eq.(\ref{la2}) in terms of the source EM field propagator resembles Schwinger's approach \cite{SchwingerMath}. In either case, the normalization by the energy of free-space fluctuations amounts to the substraction of the zero-point EM energy and bare atomic bonding energy.


\subsection{Average Lamb shift and total vacuum energy of a random medium}\label{S3C}
Finally, let us compute the average vacuum energy of a random medium, $\mathcal{F}^{V}_{avg}$ \footnote{In principle, it could be possible to include the randomness of the dipole configurations in the actual (not averaged) internal dynamics of the dipoles as considering the computation of the actual (not averaged) total vacuum energy. For this to be the case the correlation time of the density fluctuations should be much less than the typical dynamical scale, $\sim2\pi/\omega_{0}$. However, under these circumstances it is likely that the dynamics of the external d.o.f. be fast, which would induce additional complications like Doppler effects.}. In principle, since some stochastic observables can be calculated out of the optical response functions of the dielectric --$\bar{\chi}^{\omega}$, $\tilde{\alpha}(\omega)$, $\bar{\Pi}^{\omega}$, $\bar{G}^{\omega}$, $\bar{\mathcal{G}}^{\omega}$, it may seem possible to give a closed formula for $\mathcal{F}^{V}_{avg}$ which depends only on the electrical susceptibility and the renormalized polarizability. The computation is however far more complicated since the ensemble average over $\mathcal{F}^{V}_{m}$ involves highly non-linear terms in those functions. This can be expressed as
\begin{equation}
\Bigl\langle \textrm{Tr}\Bigl[\ln{[\bar{\tilde{\mathfrak{g}}}
\cdot[\bar{G}^{(0)}]^{-1}]}\Bigr]\Bigr\rangle_{avg}\neq \textrm{Tr}\Bigl[\ln{[\Bigl\langle\bar{\tilde{\mathfrak{g}}}
\cdot[\bar{G}^{(0)}]^{-1}\Bigr\rangle_{avg}]}\Bigr].\label{ineq}
\end{equation}
\indent In the first place, the ensemble average  of Eq.(\ref{eer24}) is the average Lamb shift,
\begin{eqnarray}
\mathcal{E}^{LSh}_{avg}&=&\frac{\hbar}{2\pi\rho}\int_{0}^{\infty}\textrm{d}\omega k^{2}
\Im{}\Bigl\{\int\frac{\textrm{d}^{3}q}{(2\pi)^{3}} \Bigl[2G_{\perp}^{(0)}\Pi_{\perp}+
G_{\parallel}^{(0)}\Pi_{\parallel}\Bigr]\Bigr\}\nonumber\\
&=&\frac{\hbar}{2\pi}\int_{0}^{\infty}\textrm{d}\omega k^{2}
\Im{}\{\int\frac{\textrm{d}^{3}q}{(2\pi)^{3}}\tilde{\alpha}\Bigl[2\mathcal{G}^{'}_{\perp}+
\mathcal{G}^{'}_{\parallel}\Bigr]\Bigr\}\label{eer2222441}\\
&=&\frac{\hbar}{2\pi}\int_{0}^{\infty}\textrm{d}\omega\:
\Im{}\{\tilde{\alpha}\phi\},\label{eer222244}
\end{eqnarray}
where we can identify the average density of states, $\mathcal{N}^{LSh}_{avg}=\rho\Im{}\{\frac{\tilde{\alpha}}{2\pi}\phi\}$.
As noted by Bullough in \cite{BulloughJPhysA2}, the integration of the equation analogous to that of Eq.(\ref{eer224}) for $\mathcal{F}^{V}_{avg}$,
\begin{equation}
\mathcal{F}^{V}_{avg}=\int_{0}^{\infty}\hbar\textrm{d}\omega\int_{0}^{\alpha_{0}}
\frac{\delta\alpha^{'}_{0}}{\alpha^{'}_{0}}\mathcal{N}^{LSh}_{avg},\label{nointegrable}
\end{equation}
needs of the knowledge of the functional dependence of $\bar{\chi}$ on the polarizability $\alpha_{0}$. Thus, $\mathcal{E}^{LSh}_{avg}$ can be expressed in closed form as function of $\bar{\chi}$ using Eq.(\ref{LDOSIparal}),
\begin{eqnarray}
\mathcal{E}^{LSh}_{avg}&=&\frac{\hbar}{2\pi\rho}\int_{0}^{\infty}\textrm{d}\omega k^{2}
\Im{}\Bigl\{\int\frac{\textrm{d}^{3}q}{(2\pi)^{3}} \Bigl[2\chi_{\perp}G_{\perp}^{(0)}[1+k^{2}G_{\perp}^{(0)}\chi_{\perp}]^{-1}\nonumber\\
&+&\chi_{\parallel}G_{\parallel}^{(0)}[1+k^{2}G_{\parallel}^{(0)}\chi_{\parallel}]^{-1}\Bigr]\Bigr\}.\label{eerr22224}
\end{eqnarray}
From Eq.(\ref{eerr22224}) we define transverse and longitudinal density of states per unit of momentum volume, $\mathcal{N}^{LSh}_{avg}|_{\perp,\parallel}$,
\begin{equation}
\mathcal{N}^{LSh}_{avg}|_{\perp,\parallel}=\Im{}\{(2)_{\perp}k^{2}\chi^{\omega}_{\perp,\parallel}(q)
G_{\perp,\parallel}^{(0)}[1+k^{2}G_{\perp,\parallel}^{(0)}\chi_{\perp,\parallel}^{\omega}]^{-1}\},
\end{equation}
where the prefactor $(2)_{\perp}$ applies only to transverse modes, so that
\begin{equation}
\mathcal{E}^{LSh}_{avg}=\int_{0}^{\infty}\hbar\textrm{d}\omega\int\frac{\textrm{d}^{3}q}{(2\pi)^{3}}[\mathcal{N}^{LSh}_{avg}|_{\perp}+\mathcal{N}^{LSh}_{avg}|_{\parallel}]/\rho\nonumber\\
\end{equation}
\indent However,  except for some specific  models, $\mathcal{F}^{V}_{avg}$ cannot be given in closed form. A way to go around this problem is to explode the cluster decomposition outlined previously for $\bar{\chi}$, $\bar{G}$ and $\phi$. Firstly, by taking the ensemble average over the many-body expansion of Eq.(\ref{manybody}) we can obtain a series for $\mathcal{F}^{V}_{avg}$ in multiple scattering terms. Note that this is not yet a series in $\rho$ since, at a given order $n$ of Eq.(\ref{manybody}), there are processes with $2,3,..n$ different indices. When performing upon them the ensemble average, they contribute with terms of order $\rho\alpha\phi^{(1)}...\rho\alpha\phi^{(n-1)}$ respectively. Equal index terms in Eq.(\ref{manybody}) involve a self-correlation functions $\delta^{(3)}(\vec{R}_{a}-\vec{R}_{b})$ when performing the ensemble average.\\
\indent Further, some realistic approximations can be carried out in order to obtain an expansion in powers of $\rho$. In the first place, radiative-reaction corrections in the renormalized polarizabilities which enter any diagram in $\phi^{(n)}$ can be disregarded in good approximation. Upon integration in $\omega$ both in Eq.(\ref{eer222244}) and in the average of Eq.(\ref{manybody}), they yield terms of order $\sim k_{0}r_{e}\sim \Gamma_{0}/\omega_{0}\ll1$ smaller than the computations involving bare polarizabilities, $r_{e}$ being the electron radius and $\Gamma_{0}$ being the free-space decay rate of an oscillator. Second, entangled intermediate recurrent scattering events can be ignored since they provide terms of orders $[(k_{0}\xi)^{-3}\Gamma_{0}/\omega_{0}]^{sm}\ll1$, $m\geq2$, $s=1,2,3,..,n$, smaller than the non-recurrent, non-retarded near field terms and terms of orders $[(k_{0}\xi)^{-1}\Gamma_{0}/\omega_{0}]^{sm}\ll1$, $m\geq2$, $s=1,2,3,..,n$,  smaller than the non-recurrent, retarded radiative term at any given order $n$ in $\rho$, with $\xi$ being the typical spatial correlation length, $s$ the number of scatterers repeated and $m$ the recurrent multiplicity --cf. computation of $\phi^{(1)}$ in Sec.\ref{S5} and Appendix \ref{appendyo}. With these conditions provided, we can simplify Eq.(\ref{eer222244}) and the average of Eq.(\ref{manybody}) as
\begin{eqnarray}
\mathcal{E}^{LSh}_{avg}&=&\mathcal{E}^{LSh}_{0}+\mathcal{E}^{LSh}_{sc}\nonumber\\
&\simeq&\mathcal{E}^{LSh}_{0}+
\frac{\hbar}{2\pi}\int_{0}^{\infty}\textrm{d}\omega\:
\Im{}\Bigl\{\int\frac{\textrm{d}^{3}q}{(2\pi)^{3}}\alpha_{0}\sum_{m=1}^{\infty}\phi^{(m,0)}_{\alpha_{0}}\Bigr\}\nonumber\\
&\equiv&\mathcal{E}^{LSh}_{avg}|_{\alpha_{0}},\label{laEshapprox}\\
\mathcal{F}^{V}_{avg}&\simeq&\mathcal{F}^{L}_{0}+\frac{\hbar}{2\pi}\int_{0}^{\infty}\textrm{d}\omega
\Im{}\Bigl\{\sum_{m=1}^{\infty}\frac{1}{m+1}\rho\alpha_{0}\phi^{(m,0)}_{\alpha_{0}}\Bigr\}\nonumber\\
&\equiv&\mathcal{F}^{V}_{avg}|_{\alpha_{0}},\label{manybodyapprox}
\end{eqnarray}
where the scattering Lamb shift, $\mathcal{E}^{LSh}_{sc}$, has been isolated explicitly and  $\phi^{(m,0)}_{\alpha_{0}}(\omega)$ stands for the $m$-body term of the expansion of $\phi$ in which no entangled intermediate recurrent scattering events appear (norec.) and polarizabilites are taken as bare.\\
\indent Lastly, we can write Eqs.(\ref{laEshapprox},\ref{manybodyapprox}) in function of the cluster expansion for $\bar{\chi}$. This way, we obtain a formula for $\mathcal{F}^{V}_{avg}|_{\alpha_{0}}$ which depends explicitly on (the cluster expansion of) $\bar{\chi}$, order by order in $\rho$. As for Eq.(\ref{eerr22224}) we use Eq.(\ref{LDOSIparal}) and the precedent approximation to write,
\begin{eqnarray}
\rho\alpha_{0}\phi^{(norec.)}_{\alpha_{0}}&=&k^{2}\int\frac{\textrm{d}^{3}q}{(2\pi)^{3}} \Bigl[2\chi^{norec.}_{\perp,\alpha_{0}}G_{\perp}^{(0)}[1+k^{2}G_{\perp}^{(0)}\chi^{norec.}_{\perp,\alpha_{0}}]^{-1}\nonumber\\
&+&\chi^{norec.}_{\parallel,\alpha_{0}}G_{\parallel}^{(0)}[1+k^{2}G_{\parallel}^{(0)}\chi^{norec.}_{\parallel,\alpha_{0}}]^{-1}\Bigr]\label{approxrhoalphaphi}\\
&=&\int\frac{\textrm{d}^{3}q}{(2\pi)^{3}} \sum_{m=1}(-1)^{m+1}\Bigl[2[k^{2}\chi^{norec.}_{\perp,\alpha_{0}}G_{\perp}^{(0)}]^{m}\nonumber\\
&+&[k^{2}\chi^{norec.}_{\parallel,\alpha_{0}}G_{\parallel}^{(0)}]^{m}\Bigr],
\end{eqnarray}
where $\chi^{norec.}_{\perp,\parallel,\alpha_{0}}$ are the transverse and longitudinal components of the susceptibility with  bare polarizabilities and no entangled recurrent scattering processes. Using the cluster expansion of Eq.(\ref{laXenAes}) we obtain,
\begin{eqnarray}
\mathcal{E}^{LSh}_{avg}|_{\alpha_{0}}&=&\mathcal{E}^{LSh}_{0}+\frac{\hbar}{2\pi\rho}\int_{0}^{\infty}\textrm{d}\omega
\Im{}\{\int\frac{\textrm{d}^{3}q}{(2\pi)^{3}}2[k^{2}\chi^{(2,0)}_{\perp,\alpha_{0}}G_{\perp}^{(0)}\nonumber\\&-&
(k^{2}\chi^{(1,0)}_{\perp,\alpha_{0}}G_{\perp}^{(0)})^{2}+(k^{2}\chi^{(1,0)}_{\perp,\alpha_{0}}G_{\perp}^{(0)})^{3}\nonumber\\
&+&k^{2}G_{\perp}^{(0)}\chi^{(3,0)}_{\perp,\alpha_{0}}
-2\chi^{(1,0)}_{\perp,\alpha_{0}}\chi^{(2,0)}_{\perp,\alpha_{0}}(k^{2}G_{\perp}^{(0)})^{2}+..]
\nonumber\\&+&[k^{2}\chi^{(2,0)}_{\parallel,\alpha_{0}}G_{\parallel}^{(0)}-(k^{2}\chi^{(1,0)}_{\parallel,\alpha_{0}}G_{\parallel}^{(0)})^{2}+(k^{2}\chi^{(1,0)}_{\parallel,\alpha_{0}}G_{\parallel}^{(0)})^{3}\nonumber\\
&+&k^{2}G_{\parallel}^{(0)}\chi^{(3,0)}_{\parallel,\alpha_{0}}
-2\chi^{(1,0)}_{\parallel,\alpha_{0}}\chi^{(2,0)}_{\parallel,\alpha_{0}}(k^{2}G_{\parallel}^{(0)})^{2}+..]\},\label{laEshapproxchi}\\
\mathcal{F}^{V}_{avg}|_{\alpha_{0}}&=&\mathcal{F}^{L}_{0}+\frac{\hbar}{2\pi}\int_{0}^{\infty}\textrm{d}\omega
\Im{}\Bigl\{\int\frac{\textrm{d}^{3}q}{(2\pi)^{3}}2\Bigl[\frac{1}{2}\Bigl(k^{2}\chi^{(2,0)}_{\perp,\alpha_{0}}G_{\perp}^{(0)}\nonumber\\&-&(k^{2}\chi^{(1,0)}
_{\perp,\alpha_{0}}G_{\perp}^{(0)})^{2}\Bigr)
+\frac{1}{3}\Bigl((k^{2}\chi^{(1,0)}_{\perp,\alpha_{0}}G_{\perp}^{(0)})^{3}\nonumber\\&+&k^{2}G_{\perp}^{(0)}\chi^{(3,0)}_{\perp,\alpha_{0}}
-2\chi^{(1,0)}_{\perp,\alpha_{0}}\chi^{(2,0)}_{\perp,\alpha_{0}}(k^{2}G_{\perp}^{(0)})^{2}\Bigr)+..\Bigr]
\nonumber\\&+&\Bigl[\frac{1}{2}\Bigl(k^{2}\chi^{(2,0)}_{\parallel,\alpha_{0}}G_{\parallel}^{(0)}-(k^{2}\chi^{(1,0)}_{\parallel,\alpha_{0}}G_{\parallel}^{(0)})^{2}\Bigr)
\nonumber\\&+&\frac{1}{3}\Bigl((k^{2}\chi^{(1,0)}_{\parallel,\alpha_{0}}G_{\parallel}^{(0)})^{3}+k^{2}G_{\parallel}^{(0)}\chi^{(3,0)}_{\parallel,\alpha_{0}}
\nonumber\\&-&2\chi^{(1,0)}_{\parallel,\alpha_{0}}\chi^{(2,0)}_{\parallel,\alpha_{0}}(k^{2}G_{\parallel}^{(0)})^{2}\Bigr)+..\Bigr]\Bigr\},\label{manybodyapproxchi}
\end{eqnarray}
which in turn yields an expansion in powers of $\rho e^{2}$ like that of Mclachlan \cite{McLachlan63}.
It is plain that the problem in integrating Eq.(\ref{nointegrable}) has been shifted to that of knowing $\chi_{\perp,\parallel}(q;\omega)$ as a power series of $\rho$ or, otherwise, knowing all order spatial correlation functions.

\section{Approximations to the total vacuum energy of a random medium}\label{S4}

\subsection{The quasi-crystalline approximation}

\begin{figure}[h]
\includegraphics[height=3.4cm,width=8.2cm,clip]{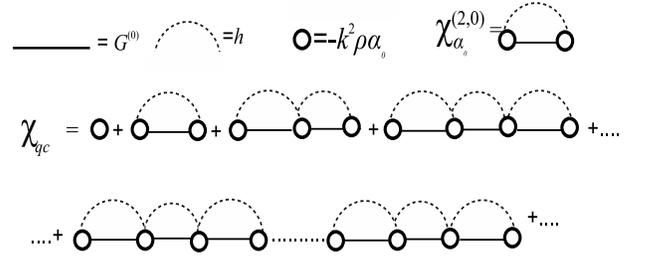}
\caption{Diagrammatic representation of Eq.(\ref{laE}).}\label{figb2}
\end{figure}

In previous works --cf. \cite{Obada,BulloughJPhysA2,PRAvc}, it has been reported closed formulas for $\mathcal{F}^{V}_{avg}$ as functions of $\chi_{\perp,\parallel}$. However, they are model dependent. Nonetheless, they are useful for estimating orders of magnitude. In particular, that in \cite{PRAvc} corresponds to the so called quasi-crystalline approximation (qc). According to this approximation the only relevant correlation function is the two-body one, $h(r)$, and self-correlations are ignored. In this approximation, the series of $\bar{\chi}(q)$ becomes geometrical and the only quantity to be computed is $\bar{\chi}^{(2,0)}(q)$,
\begin{equation}
\chi^{(2,0)}_{p,\alpha_{0}}(q)=\frac{-k^{2}\alpha^{2}_{0}\rho^{2}}{(1+\delta_{\perp}^{p})}
\int\textrm{d}^{3}r e^{i\vec{q}\cdot\vec{r}}h(r)\textrm{Tr}
\{\bar{G}^{(0)}(\vec{r})\cdot\bar{\textrm{P}}_{p}(\hat{q})\},\label{LLFF}
\end{equation}
with $p=\perp,\parallel$. The geometrical series in Fig.\ref{figb2} is
\begin{equation}\label{laE}
\chi_{\perp,\parallel}^{qc}(q;\omega)=\frac{\rho\alpha_{0}}{1-\chi_{\perp,\parallel,\alpha_{0}}^{(2,0)}(q;\omega)/\rho\alpha_{0}},
\end{equation}
and so,
\begin{equation}
\mathcal{G}_{\perp,\parallel}^{qc}(q,\omega)=G_{\perp,\parallel}^{(0)}[1+\rho\alpha_{0}(k^{2}G_{\perp,\parallel}^{(0)}-
\chi^{(2,0)}_{\perp,\parallel,\alpha_{0}}/\rho^{2}\alpha^{2}_{0})]^{-1},
\end{equation}
and
\begin{eqnarray}
\mathcal{F}^{V}_{qc}&=&\frac{-\hbar}{2\pi}\int_{0}^{\infty}\textrm{d}\omega
\Im{}\Bigl\{\int\frac{\textrm{d}^{3}q}{(2\pi)^{3}}\nonumber\\&\times& \ln{} \Bigl[[\mathcal{G}_{\perp,\alpha_{0}}^{qc}]^{2}\mathcal{G}_{\parallel,\alpha_{0}}^{qc}[G_{\perp}^{(0)}]^{-2}[G_{\perp}^{(0)}]^{-1}\Bigr]\Bigr\}\label{laFqc}\\
&=&\frac{\hbar}{2\pi}\int_{0}^{\infty}\textrm{d}\omega\Im{}\Bigl\{\int\frac{\textrm{d}^{3}q}{(2\pi)^{3}}\nonumber\\&\times&\Bigl[ 2\ln{} [1+\rho\alpha_{0}(k^{2}G_{\perp}^{(0)}-\chi^{(2,0)}_{\perp,\alpha_{0}}/\rho^{2}\alpha^{2}_{0})]\nonumber\\
&+&\ln{} [1+\rho\alpha_{0}(k^{2}G_{\parallel}^{(0)}-
\chi^{(2,0)}_{\parallel,\alpha_{0}}/\rho^{2}\alpha^{2}_{0})]\Bigr]\Bigr\},\label{Fqc}
\end{eqnarray}
where by ignoring radiation-reaction it holds that $\tilde{\mathcal{G}}_{\perp,\parallel}^{qc}=\mathcal{G}_{\perp,\parallel}^{qc}$.
It is because of the aforementioned analogy between the formulas for a specific configuration of dipoles --cf. a cubic lattice \cite{Obada}-- and those of the quasicrystalline approximation, that the inequality expressed in Eq.(\ref{ineq}) turns into an equality in this case. It is also possible to write Eq.(\ref{Fqc}) as
\begin{eqnarray}
\mathcal{F}^{V}_{qc}&=&-\hbar\Im{}\Bigl\{\int_{0}^{\infty}\frac{\textrm{d}\omega}{2\pi} \int\frac{\textrm{d}^{3}q}{(2\pi)^{3}}2\ln{}[\chi^{qc}_{\perp}G^{qc}_{\perp}]
+\ln{}[\chi^{qc}_{\parallel}G^{qc}_{\parallel}]\Bigr\}\label{bulky}\\
&+&\hbar\Im{}\Bigl\{\int_{0}^{\infty}\frac{\textrm{d}\omega}{2\pi} \int\frac{\textrm{d}^{3}q}{(2\pi)^{3}}2\ln{}\Bigl[[G^{(0)}_{\perp}]\Bigr]+\ln{}\Bigl[[G^{(0)}_{\parallel}]\Bigr]\Bigr\}\label{lacagada}\\
&+&3\hbar\Im{}\Bigl\{\int_{0}^{\infty}\frac{\textrm{d}\omega}{2\pi} \int\frac{\textrm{d}^{3}q}{(2\pi)^{3}}\ln{[\alpha_{0}]}\Bigr\}.\label{Flamb}
\end{eqnarray}
Making the identification $\int\frac{\textrm{d}^{3}q}{(2\pi)^{3}}=\rho$ in Eq.(\ref{Flamb}) \footnote{This equivalence can be proved passing to the continuum from a lattice of dipoles whose cell volume is $\rho^{-1}$. It acts as a regulator.} we read that the bare atomic bonding energy and the EM zero-point energy enter as substraction terms. This justifies the interpretation of $\mathcal{F}^{V}_{qc}$ in \cite{Obada}. That is, $\mathcal{F}^{V}_{qc}$ takes account of the zero-point energy of bare EM modes in Eq.(\ref{lacagada}), and the atomic bonding energy of Eq.(\ref{Flamb}) and substitutes them with the binding energy of the coupled system of Eq.(\ref{bulky}).\\
\indent Further, by considering the continuum limit of the effective dielectric constant, $q\xi\ll1$, we will see that $\mathcal{F}^{V}_{qc}(q\xi\ll1)$ can be written as a function of the refractive index. We will discuss the accuracy of this approximation.\\
\indent  Although the expression obtained in \cite{PRAvc} for $\mathcal{F}^{V}_{avg}$ coincides with that of Eq.(\ref{laFqc}), the derivation there was erroneous and hence its validity restricts to the quasicrystalline approximation. The reasoning followed in \cite{PRAvc} was that the average energy could be obtained by extending appropriately Schwinger's approach on an effective continuum medium \cite{Schwinger4095} to a molecular dielectric. Thus, the steps followed by Schwinger \emph{et al.} in \cite{deRaad} and Schwinger in \cite{Schwinger4095,Schwinger2105} were mimicked in \cite{PRAvc} but for the fact that the effective bulk propagator was replaced by the average source field propagator of a molecular dielectric. The reason for doing this was the constatation in \cite{PRAvc} that the effective bulk propagator used in \cite{deRaad,Schwinger4095,Schwinger2105} is not the one which enters the formula for the Lamb shift  in Eq.(\ref{eer2222441}). Accidentally the approach in \cite{PRAvc} turns the inequality of Eq.(\ref{ineq}) into an equality, which is generally incorrect.

\subsection{The Schwinger bulk energy of an effective medium}\label{S4B}
The reason why the approach of Schwinger \emph{et al.} in \cite{deRaad} and Schwinger in \cite{Schwinger2105} does not yield the total binding energy of a molecular dielectric is that the variation of the energy shift considered there is not appropriate to this aim \footnote{Note that this does not imply by any means that the generic Schwinger's source theory be inapplicable to this aim as shown explicitly for the case of a specific configuration of dipoles, Eq.(\ref{la2}) --cf. \cite{JaffePRD}.}. The EM effective action of Schwinger \emph{et al.} \cite{deRaad} is that of a piecewise homogeneous medium of effective electrical susceptibility $\chi_{eff}^{\omega}(\vec{r})$ with interaction Hamiltonian,
\begin{equation}
H^{eff}_{int}=-\int\textrm{d}^{3}r\vec{P}_{eff}(\vec{r},t)\cdot\vec{E}^{\perp}_{eff}(\vec{r},t).\label{effc}
\end{equation}
The corresponding energy shift is
\begin{equation}\label{backreact}
\mathcal{E}_{Sch.}=-\Re{\Bigr\{\int\textrm{d}^{3}r\textrm{d}\:\langle\Omega_{Sch.}|\hat{\vec{\mathbf{P}}}_{eff}(\vec{r},t)
\cdot\hat{\vec{\mathbf{E}}}^{\perp}_{eff}(\vec{r},t)|\Omega_{Sch.}\rangle\Bigl\}},
\end{equation}
where $\vec{E}^{\perp}_{eff}$ is the transverse  effective field generated by an effective polarization density source $\vec{P}_{eff}$ through the relation,
\begin{equation}\label{Esource}
\vec{E}^{\perp}_{eff}(\vec{r};\omega)=k^{2}\epsilon_{0}^{-1}\int\textrm{d}^{3}r'\bar{G}^{eff}_{\perp}(\vec{r},\vec{r}';\omega)\cdot\vec{P}^{\omega}_{eff}(\vec{r}').
\end{equation}
The response function of the source field  is the transverse effective  bulk (Dyson) propagator,
$G^{eff}_{\perp}(q;\omega)=(\epsilon_{eff}k^{2}-q^{2})^{-1}$. Adapting the previous nomenclature, $G^{eff}_{\perp}(q;\omega)$ characterizes the EM field
fluctuations of the 'Schwinger vacuum' (Sch.). By varying the action of the effective medium, Schwinger \emph{et al.} \cite{deRaad} inferred the identification
of the effective polarization source quadratic fluctuations,
\begin{eqnarray}
\int&\textrm{d}&t\exp{[i\omega t]}i(\epsilon_{0}\hbar)^{-1}\Theta(t)\nonumber\\&\times&\langle\Omega_{Sch.}|\hat{\vec{\mathbf{P}}}_{eff}(\vec{r},0)\otimes\hat{\vec{\mathbf{P}}}_{eff}(\vec{r}',t)|\Omega_{Sch.}\rangle
=\delta\bar{\chi}^{\omega}_{eff}\delta^{(3)}(\vec{r}-\vec{r}').\nonumber
\end{eqnarray}
Inserting the above expression together with Eq.(\ref{Esource}) into Eq.(\ref{backreact}) we get \cite{Schwinger2105},
\begin{eqnarray}
\delta_{\chi'}\mathcal{E}_{Sch.}&=&\frac{\hbar}{2\pi\rho}\int_{0}^{\infty}\textrm{d}\omega k^{2}
\Im{}\Bigl\{\int\frac{\textrm{d}^{3}q}{(2\pi)^{3}}2\delta\chi^{'\omega}_{eff}G^{eff}_{\perp}\Bigr\}\nonumber\\
&=&\frac{\hbar}{2\pi\rho}\int_{0}^{\infty}\textrm{d}\omega k^{2}
\Im{}\Bigl\{\int\frac{\textrm{d}^{3}q}{(2\pi)^{3}}2\delta\chi^{'\omega}_{eff}
G_{\perp}^{(0)}\nonumber\\&\times&[1+k^{2}G_{\perp}^{(0)}\chi_{eff}^{'\omega}]^{-1}\Bigr\}.\label{Scheer22224}
\end{eqnarray}
The functional integration immediately yields
\begin{equation}
\mathcal{F}^{V}_{Sch.}=\frac{-\hbar}{2\pi}\int_{0}^{\infty}\textrm{d}\omega
\Im{}\{\int\frac{\textrm{d}^{3}q}{(2\pi)^{3}} \ln{} \bigl[[G^{eff}_{\perp}]^{2}[G_{\perp}^{(0)}]^{-2}\bigr]\},\label{FSch}
\end{equation}
which contains only the fluctuations of the transverse effective bulk propagator.
\subsubsection{Relation with the microscopic approach}
In order to understand the link with the microscopic computation we first notice the similarity between Eq.(\ref{FSch}) and Eq.(\ref{eerr22224}) for the microscopic computation of $\mathcal{E}^{LSh}_{avg}$. Ignoring the fact that Schwinger neglected the contribution of longitudinal modes, Eq.(\ref{FSch}) can be written also as
\begin{equation}
\mathcal{F}^{V}_{Sch.}=\int\hbar\textrm{d}\omega\int\frac{\textrm{d}^{3}q}{(2\pi)^{3}}\int_{0}^{\chi^{\omega}_{eff}}\frac{\delta\chi^{'\omega}_{eff}}{\chi^{'\omega}_{eff}}
\mathcal{N}^{LSh}_{avg}|^{eff}_{\perp},\label{IntegraSch}
\end{equation}
where $\mathcal{N}^{LSh}_{avg}|^{eff}_{\perp}$ is the average density of states restricted to transverse modes in the continuum.
This implies that, effectively, in the approach of Schwinger \emph{et al.}  the microscopical interaction Hamiltonian of Eq.(\ref{Hamil}) with dipole moment operator $e\hat{\vec{\mathbf{r}}}$ and coupling constant $e$ is substituted with the effective Hamiltonian of Eq.(\ref{effc}), which couples the transverse effective EM field to an effective polarization density operator $\hat{\vec{\mathbf{P}}}_{eff}$ with some effective coupling constant proportional to $\chi^{1/2}_{eff}$. Because $\chi_{eff}$ is made of the integration of dipole-dipole interactions, the energy of those interactions is disregarded in the approach of Schwinger \emph{et al.} \footnote{Analogously, in assuming a given coupling constant and a bare resonant frequency, the computation of the internal binding energy of the atomic dipoles prior to the coupling to radiation is also disregarded in our approach.}. This means that Schwinger \emph{et al.} approach is not appropriate to study phenomena driven by the variation of the energies of internal d.o.f. This is the case of Lamb shifts \cite{Milonnietal}, van der Waals forces and phase transitions in liquid crystals \cite{ZhengPalffy} and colloidal suspensions \cite{Verweybook} in which the relevant variations in energy are those w.r.t. that of the molecular constituents infinitely far apart. However, it is appropriate for the treatment of problems in which the internal properties of the objects are integrated in the definition of their permittivities and remain unaffected during the phenomena under study. This is the case of the Casimir forces between macroscopic dielectrics separated by macroscopic distances \cite{deRaad} in which the relevant variations in energy are those w.r.t. that of the macroscopic objects infinitely far apart. Nonetheless, the reason why $\mathcal{E}^{LSh}_{0}$ can be computed out of $\mathcal{F}^{V}_{Sch.}$ \cite{Schaden} is that, at leading order in $\rho$, $\bar{\chi}(\vec{r})\simeq\rho\alpha_{0}\bar{\mathbb{I}}\delta^{(3)}(\vec{r})$ and so $\mathcal{F}^{V}_{0}=\mathcal{F}^{V}_{Sch.}$ at $\mathcal{O}(\rho)$.\\
\indent The original suggestion of Schwinger to explain sonoluminescence represents a sort of intermediate problem \cite{Schwinger2105}. Basically, when a bubble embedded in water collapses, molecules of water fill in the void and he conjectured  that the light emitted in this process carries the excess of EM vacuum energy stored in the void w.r.t. the energy of the homogenous aqueous medium. Evidently the molecules of water which fill the void form part of the initial aqueous medium surrounding the bubble. Therefore, leaving aside the validity of further dynamical approximations -- cf. \cite{MiltonarXiv}, the total volume filled by the water is greater after the collapse, the density of water is less and so the dielectric constant must vary, at least locally, in the region within and around the primordial bubble. As a consequence, it is reasonable to think that, if the time scale of light emission is less than the typical homogenization time, the molecules of water within the volume originally occupied by the bubble will have different spatial disposition to that of the homogenous surrounding medium. Therefore, their internal energy will be different to that of the homogeneous phase and the problem involves not only a variation of energy between macroscopic objects (aqueous media with and without a bubble) but also a difference in the internal energy of the molecules which were initially in a homogenous disposition and fill in the void afterwards in a different arrangement. On the contrary, if the homogenization is reached prior to emission and, in good approximation, the permittivity of the filled bubble is identical to that of the surrounding medium, Schwinger's approach might be a good approximation.\\
\subsubsection{Beyond the continuum approximation}
The result of Schwinger in the continuum yields only the energy associated to radiative modes which depends solely on the effective refractive index, $n(\omega)=\sqrt{1+\chi^{\omega}_{eff}}
$, as
\begin{equation}\label{Fschn}
\mathcal{F}_{Sch.}^{V}\simeq\frac{\hbar}{6\pi^{2}c^{3}}\Re{\Bigl\{\int_{0}^{\infty}\textrm{d}\omega\:\omega^{3}[1-n^{3}]\Bigr\}},
\end{equation}
and the Schwinger energy shift is
\begin{equation}\label{LSHSCH}
\mathcal{E}_{Sch.}\simeq \frac{-\hbar}{4\pi^{2}c^{3}\rho}\Re{}\bigl\{\int_{0}^{\infty}\textrm{d}\omega\:\omega^{3}n\chi^{\omega}_{eff}\bigr\}.
\end{equation}
In Sec.\ref{DiscLSch} we compare in detail the above results with the actual microscopical calculation.
Schwinger argued \cite{Schwinger2105} that these results in the continuum could be extended to incorporate spatial dispersion in the electrical susceptibility. In doing so we obtain an extended (ext) Schwinger transverse bulk energy, $\mathcal{F}^{V}_{Sch.}|^{ext}$. However, even in this case the discrepancy between the actual $\mathcal{F}^{V}_{avg}$ and $\mathcal{F}^{V}_{Sch.}|^{ext}$ appears already at order $\rho^{2}$. For the sake of simplicity we restrict ourselves to the approximations used in  Eq.(\ref{manybodyapproxchi}) and find,
\begin{eqnarray}
\mathcal{F}^{V}_{avg}|_{\alpha_{0}}&\simeq&\mathcal{F}^{V}_{avg}|^{\parallel}_{\alpha_{0}}+\mathcal{F}^{V}_{Sch.}|^{ext}_{\alpha_{0}}+\frac{\hbar}{2\pi}\int_{0}^{\infty}\textrm{d}\omega\:
\Im{}\Bigl\{\int\frac{\textrm{d}^{3}q}{(2\pi)^{3}}2\nonumber\\
&\times&\Bigl[-\frac{1}{2}k^{2}\chi^{(2,0)}_{\perp,\alpha_{0}}G_{\perp}^{(0)}\label{discrep}
\nonumber\\&-&\frac{2}{3}k^{2}G_{\perp}^{(0)}\chi^{(3,0)}_{\perp,\alpha_{0}}
-\frac{2}{3}\chi^{(1,0)}_{\perp,\alpha_{0}}\chi^{(2,0)}_{\perp,\alpha_{0}}(k^{2}G_{\perp}^{(0)})^{2}\nonumber\\
&-&(k^{2}\chi^{(1,0)}_{\perp,\alpha_{0}}G_{\perp}^{(0)})^{3}+...\Bigr],
\end{eqnarray}
where $\mathcal{F}^{V}_{avg}|^{\parallel}_{\alpha_{0}}$ contains the longitudinal modes whose contribution includes radiative energy --cf. computation of  $\mathcal{F}^{V}_{\mathcal{O}(\rho^{2})}$ in Sec.\ref{S5A}.\\
\indent For the sake of completeness we mention that the approach of Abrikosov, Dzyaloshinskii, Gorkov, Lifshitz and Piaevskii (ADGLP) \cite{Dzy,Abriko} is based on a semi-phenomenological prescription for the vacuum energy which accounts for the restriction of the EM fluctuations to those which contain a strictly local \emph{polarization operator}, $\sim\chi_{eff}\bar{\mathbb{I}}\delta^{(3)}(\vec{r}-\vec{r}')$. Hence, its content is similar to that of Schwinger's in the continuum but for the fact that it includes the contribution of long-wavelength bulk longitudinal modes,
\begin{equation}\label{ADGLP}
\mathcal{F}_{ADGLP}=-\hbar\Im{}\Bigl\{\int_{0}^{\infty}\frac{\textrm{d}\omega}{2\pi}
\int\frac{\textrm{d}^{3}q}{(2\pi)^{3}}\ln{}\Bigl[\frac{(k^{2}-q^{2})^{2}}{\epsilon_{eff}(\epsilon_{eff} k^{2}-q^{2})^{2}}\Bigr]\Bigr\},
\end{equation}
where $\epsilon_{eff}$ is the effective dielectric constant. A critical analysis of it has been carried out by Bullough in \cite{BulloughJPhysA5}.

\section{The leading order Lamb shift and vacuum energy in a random medium}\label{S5}
At leading order, the free-space Lamb shift and Lamb energy are additive. Also, at leading order in $\rho$, $\mathcal{F}^{L}_{0}\simeq\rho\mathcal{E}^{LSh}_{0}$. The actual computation of $\mathcal{F}^{L}_{0},\mathcal{E}^{LSh}_{0}$ has been carried out by a number of authors \cite{BulloughThomson,BulloughJPhysA2,Power,MilonniScripta2},
\begin{eqnarray}
\mathcal{F}^{L}_{0}&=&\hbar\rho\Im{}\Bigl\{\int_{0}^{\infty}\frac{\textrm{d}\omega}{2\pi}
\ln{[1-i\frac{\omega^{3}}{2\pi c^{3}}\alpha_{0}]}\Bigr\}\label{flambo0}\\
&\simeq&\rho\mathcal{E}^{LSh}_{0,\alpha_{0}}=-\hbar\rho\Im{}\Bigl\{\int_{0}^{\infty}\frac{\textrm{d}\omega}{(2\pi)^{2}}i\frac{\omega^{3}}{c^{3}}\alpha\Bigr\}
\\&\simeq&-\hbar\rho\Re{}\Bigl\{\int_{0}^{\infty}\frac{\textrm{d}\omega}{(2\pi)^{2}}\frac{\omega^{3}}{c^{3}}\alpha_{0}\Bigr\}.\label{flambo}
\end{eqnarray}
Upon substraction of the divergent free-electron self-energy \cite{MilonniScripta2}, $\mathcal{E}_{e}=\frac{e^{2}\hbar}{\pi m_{e}c^{3}}\int \textrm{d}\omega\:\omega$, the above integral presents a UV divergence which needs to be regularized. In the non-relativistic approximation, the 'natural choice' for the UV cut-off $\Lambda$ is that of the Compton wavelength of the electron such that $\Lambda_{C}^{e}=2\pi c\lambda_{C}^{e}=m_{e}c^{2}/\hbar$. For a single oscillator with  $\mu^{2}=\frac{e^{2}\hbar}{2m_{e}\omega_{0}}$,
\begin{eqnarray}
\mathcal{F}^{L}_{0}&\simeq&\frac{\rho}{3\pi}\alpha_{f}\frac{\hbar\omega_{0}}{m_{e}c^{2}}
\ln\Bigl[\frac{m_{e}c^{2}}{\hbar\omega_{0}}\Bigr]\hbar\omega_{0}\label{lfo}\\
&\simeq&\frac{\rho}{2\pi}\ln\Bigl[\frac{m_{e}c^{2}}{\hbar\omega_{0}}\Bigr]\hbar\Gamma_{0}\label{lfoalter}
\end{eqnarray}
where $r_{e}=\frac{e^{2}}{4\pi\epsilon_{0}m_{e}c^{2}}$ is the electron radius and $\alpha_{f}=\frac{e^{2}}{4\pi\epsilon_{0}\hbar c}$ is the fine-structure constant. Eq.(\ref{lfo}) equals Bethe's result when expressing the atomic energy in terms of the velocity of a bounded electron \cite{BulloughJPhysA2,Bethe} \footnote{Also, a similar expression to that in Eq.(\ref{lfo}) was obtained by Welton
\cite{Welton} from the variation of the non-relativistic Compton scattering cross-section due to the position fluctuations of the electron.}.
Should have we computed Eq.(\ref{flambo0}) and set the wavelength cut-off at the electron radius, $\Lambda_{r}^{e}=c/r_{e}$, we would have obtained
the classical result of Dowling \cite{Dowling} by considering the singularity around this value. However, this is inconsistent with the non-relativistic
approximation. The need of a cut-off just reflects our lack of knowledge both of the internal structure of the dipoles and of the manner the EM field
 couples to the internal d.o.f.\\
\indent At higher orders the Lamb-shift is non-additive. At $\mathcal{O}(\rho)$ we have,
\begin{equation}
\mathcal{E}^{LSh}_{\mathcal{O}(\rho)}=\hbar\rho\Im{}\Bigl\{\int_{0}^{\infty}\frac{\textrm{d}\omega}{2\pi}
\:\alpha\phi^{(1)}_{\alpha}(\omega)\Bigr\},\label{rho2}
\end{equation}
where the subscript $\alpha$ means that the renromalized  porlarizabilities in $\phi^{(1)}(\omega)$ must be replaced with free-space ones in order to keep the order $\rho$ of $\mathcal{E}^{LSh}_{\mathcal{O}(\rho)}$. At this order, the corresponding $\phi$-factors read from Fig.\ref{fig1}$(d)$,
\begin{equation}\label{varphi1}
\phi_{\alpha\perp,\parallel}^{(1)}=\int\frac{\textrm{d}^{3}q}{(2\pi)^{3}}
\Bigl[-\rho\alpha[G_{\perp,\parallel}^{(0)}]^{2}+\chi_{\alpha\perp,\parallel}^{(2)}G_{\perp,\parallel}^{(0)}/(\rho\alpha)\Bigr].
\end{equation}
From the above equation we can write, in function of LFFs,
\begin{eqnarray}
\mathcal{E}^{LSh}_{\mathcal{O}(\rho)}&=&\hbar\rho\Im{}\Bigl\{\int_{0}^{\infty}\frac{\textrm{d}\omega}{2\pi}\alpha^{2}\int\frac{\textrm{d}^{3}q}{(2\pi)^{3}}
\Bigl[\chi_{\alpha\parallel}^{(2)}(q)/(\rho\alpha)^{2}-1\nonumber\\&+&2k^{2}G_{\perp}^{(0)}(q)[\chi_{\alpha\perp}^{(2)}(q)/(\rho\alpha)^{2}
-k^{2}G_{\perp}^{(0)}(q)]\Bigr]\Bigr\}.\label{rho2p}
\end{eqnarray}
This is to show how LFFs enter the Lamb shift at order $\rho$.
Related to this fact, the authors of \cite{WelshTomas} have computed the effect of LFFs on the van der Waals forces on an only dipole in an Onsager (real) cavity using Eq.(\ref{rho2}).\\
\indent At leading order in $\alpha_{0}$, it is plain from Eqs.(\ref{laEshapprox},\ref{manybodyapprox}) that $\mathcal{F}^{V}_{\mathcal{O}(\rho^{2})}|_{\alpha_{0}}=\rho\mathcal{E}^{LSh}_{\mathcal{O}(\rho)}|_{\alpha_{0}}/2$. From those equations the same simple relation holds at higher orders, $\mathcal{F}^{V}_{\mathcal{O}(\rho^{m})}|_{\alpha_{0}}=\frac{\rho}{m}\mathcal{E}^{LSh}_{\mathcal{O}(\rho^{m-1})}|_{\alpha_{0}}$. This allows for an expansion of both  $\mathcal{F}^{V}_{avg}|_{\alpha_{0}}$ and $\mathcal{E}^{LSh}_{avg}|_{\alpha_{0}}$ in $m$-body terms of order $(e^{2}\rho)^{m}$. However, beyond the approximations used there, no simple relation exists since any given order $\rho^{m}$ contains higher powers of $e^{2}$ due to entangled recurrent scattering and additional radiative corrections.

\subsection{Computation in the hard-sphere model}\label{S5A}
\indent Except in free space, the actual computation of $\mathcal{E}^{LSh}_{avg}$ and $\mathcal{F}^{V}_{avg}$  is model dependent.
Nonetheless, generic results can be obtained within the simplest analytical model. Let $h(r-\xi)$ be the two-point correlation function, $\xi$ being the correlation length. Generally, $h(r-\xi)$ can be modeled by the addition of three terms,
\begin{equation}\label{darriba}
h(r-\xi)\simeq h^{ex.}(r-\xi)+\rho^{-1}\delta^{(3)}(\vec{r})+h^{ovd}(r-\xi).
\end{equation}
In this equation, $h^{ex.}(r-\xi)$ accounts for the exclusion volume around each dipole and its precise form depends on the interaction potential between pairs of scatterers. It tends to -1 for $r\lesssim\xi$ and to zero for $r\gtrsim\xi$. Usual forms are those of a Lennard-Jones potential and a hard-sphere potential. $h^{ovd}(r-\xi)$ takes account of the overdensity of first neighbors around a given dipole. It might be relevant for high-ordered media. The three-dimensional delta function stands for the self-correlation. $h^{ex.}$ and the self-correlation functions are inherent in any molecular dielectric.\\
\indent In order to make contact with previous approaches we further demand the existence of an effective medium for the frequency range of interest. That implies $\zeta\equiv k\xi\ll1$ for $\omega\leqslant\omega_{0}$. For the sake of simplicity we will neglect in first approximation both the overdensity and the self-correlation terms in $h(r-\xi)$. We will show \emph{a posteriori} that the latter is plainly justified while the former implies slight modifications in numerical prefactors.\\
\indent Without much loss of generality we will take a hard-sphere (hs) exclusion volume correlation function, $h^{ex.}(r-\xi)=h^{hs}(r-\xi)=-\Theta(r-\xi)$, which derives from the potential $U(\vec{R}^{1},..,\vec{R}^{N})\rightarrow\infty$ if $|\vec{R}^{i}-\vec{R}^{j}|\leq\xi$, $i\neq j$, and $0$ otherwise. The numerical factors of the calculations which involve near field modes depend on the precise profile of $h^{ex.}(r-\xi)$. On the contrary, radiative propagating  modes are model-independent.\\
\indent The computation of $\phi^{(1,0)}$ in spatial coordinates is easier than that of Eq.(\ref{varphi1}) in Fourier space,
\begin{eqnarray}
\phi^{(1,0)}_{\alpha,hs}&=&\textrm{Tr}\Bigl\{\int\textrm{d}^{3}r
\:\bar{G}^{(0)}(\vec{r})(-k^{4}\rho\alpha)
\bar{G}^{(0)}(\vec{r})\nonumber\\&\times&[1-\Theta(r-\xi)]\Bigr\}\nonumber\\
&=&\frac{-k^{3}}{2\pi}\rho\alpha e^{2i\zeta}
\Bigl[\frac{1}{\zeta^{3}}-\frac{2}{\zeta^{2}}i-\frac{1}{\zeta}+\frac{1}{2}i\Bigr]\label{gammatot}\\
&\simeq&\frac{-k^{3}}{2\pi}\rho\alpha\Bigl[\frac{1}{\zeta^{3}}+\frac{1}{\zeta}+\frac{7}{6}i-\zeta\Bigr],\quad \zeta<1.\label{gammatotp}
\end{eqnarray}
The decomposition into transverse and longitudinal components is given in Appendix \ref{appendyo}.
Inserting Eq.(\ref{gammatot}) into Eq.(\ref{rho2}) and integrating in $\omega$ we obtain,
\begin{equation}\label{FL20}
\mathcal{E}^{LSh}_{\mathcal{O}(\rho)}|^{norec.}_{hs}\simeq
\frac{-\rho\mu^{2}}{12\epsilon_{0}}\frac{\Gamma_{0}}{\omega_{0}}\Bigl[(\zeta_{0}^{-3}-\zeta^{-1}_{0})+\frac{14}{3\pi}(5/6-\gamma_{\mathcal{E}}-\ln{[2\zeta_{0}]})
\Bigr],
\end{equation}
where $\gamma_{\mathcal{E}}$ is the Euler constant and, as for the computation of $\mathcal{F}_{0}^{L}$, the cut of the integrand at $\Lambda_{r}^{e}$ has been neglected. For simplicity, the integral has been expanded in powers of $\zeta_{0}=k_{0}\xi\ll1$ up to order zero, being the leading order term that of London's potential \cite{Nijboer}. In contrast to $\mathcal{F}^{L}_{0}$, the oscillating factor $e^{2i\zeta}$ serves a natural UV cut-off at $\Lambda\simeq c/2\xi$ \footnote{Making use of the property $\int_{0}^{\infty}\textrm{d}\omega\Im{\{\phi(\omega)\}}=\int_{0}^{\infty}\textrm{d}\omega\phi(i\omega)$ \cite{Wylie,Agarwal2}, all the integrals in $\omega$ are performed with the change of variables $u=-i\omega$. Since our computation refers to the ground state, resonant terms are absent.}.\\
\indent In the last equations the superscript $(1,0)$ signals that recurrent scattering terms have not been included. As advanced in Sec.\ref{S3C}, the condition $\zeta_{0}^{3}\gg\Gamma_{0}/\omega_{0}\sim k_{0}r_{e}$, implicit also in Eq.(\ref{FL20}), suffices to guarantee that recurrent scattering can be neglected in good approximation. In Eq.(\ref{A4}) of Appendix \ref{appendy} we give the expression for the complete series of recurrent scattering diagrams which amount to $\phi^{(1)}_{\alpha}$. Its expansion in powers of $\alpha$ and further integration in $\omega$ yields a series of the form,
\begin{equation}\label{expanV}
\mathcal{E}^{LSh}_{\mathcal{O}(\rho)}|_{hs}=\rho\epsilon_{0}^{-1}\mu^{2}\frac{\Gamma_{0}}{\omega_{0}}\sum_{m=0}f^{LSh}_{m}(\zeta_{0})(\Gamma_{0}/\omega_{0})^{2m}.
\end{equation}
The functions $f^{LSh}_{m}(\zeta_{0})$ contain both negative and positive powers of $\zeta_{0}$ together with terms proportional to $\ln{[(2(m+1)\zeta_{0}]}$.
Each order $m$ presents a wavelength cut-off at $4\pi(m+1)\xi$. For $m=0$, $f^{LSh}_{0}(\zeta_{0})$ is readily identifiable from Eq.(\ref{FL20}). The leading order term of a generic function $f^{LSh}_{m}(\zeta_{0})$ is of the order of $\zeta_{0}^{-3(2m+1)}$. Therefore, $[f^{LSh}_{m+1}(\Gamma_{0}/\omega_{0})^{2(m+1)}]/[f^{LSh}_{m}(\Gamma_{0}/\omega_{0})^{2m}]\sim(\frac{\Gamma_{0}/\omega_{0}}{\zeta_{0}^{3}})^{2}$ and the convergence of the series is guaranteed by the aforementioned inequality, $\zeta_{0}^{3}\gg\Gamma_{0}/\omega_{0}$. At the same time, this makes the neglect of recurrent scattering terms a good approximation.\\
\indent Because each order in recurrent scattering carries an additional factor $\alpha_{0}^{2}$, the integration of $\mathcal{E}^{LSh}_{\mathcal{O}(\rho)}|_{hs}$ yields for non-recurrent and recurrent terms respectively,
\begin{eqnarray}
\mathcal{F}^{V}_{\mathcal{O}(\rho^{2})}|^{norec.}_{hs}&\simeq&
\frac{-\rho\mu^{2}}{24\epsilon_{0}}\frac{\Gamma_{0}}{\omega_{0}}\Bigl[(\zeta_{0}^{-3}-\zeta^{-1}_{0})+\frac{14}{3\pi}(5/6-\gamma_{\mathcal{E}}\nonumber\\
&-&\ln{[2\zeta_{0}]})\Bigr],\label{100}
\end{eqnarray}
\begin{equation}
\mathcal{F}^{V}_{\mathcal{O}(\rho^{2})}|_{hs}=\rho^{2}\epsilon_{0}^{-1}\mu^{2}\frac{\Gamma_{0}}{\omega_{0}}\sum_{m=0}\frac{1}{m+2}f^{LSh}_{m}(\zeta_{0})
(\Gamma_{0}/\omega_{0})^{2m}.\label{lu}
\end{equation}
\indent The numerical constants in Eq.(\ref{FL20}) depend on the specific profile of $h^{ex.}$. Nonetheless, the order and the sign of the terms are generic. Hence, the script $hs$ can be dropped from Eqs.(\ref{expanV},\ref{lu}) since those expansions are not constrained to any particular model. As a matter of fact, the addition to Eq.(\ref{gammatot}) of the terms in Eq.(\ref{gammaovd}) which account for the overdensity two-point correlation function, $h^{ovd}(r)=C\xi\delta^{(1)}(r-\xi)$, just modifies slightly the numerical prefactors of the terms in Eqs.(\ref{FL20},\ref{100}), but neither their order nor their sign.\\
\indent It is worth mentioning that, although at a given order $\rho^{n}$ the leading contribution comes from non-recurrent terms, there are recurrent terms in orders $\rho^{s}$, $0<s<n$, which are of order $(\rho\xi^{3})^{-(n-s)}\geq1$ greater than the leading order non-recurrent term of the order $\rho^{n}$.

\subsection{Discrepancy between $\mathcal{F}^{V}_{avg}|_{rad}$ and $\mathcal{F}^{V}_{Sch.}$}\label{DiscLSch}
\begin{figure}[h]
\includegraphics[height=4.1cm,width=8.4cm,clip]{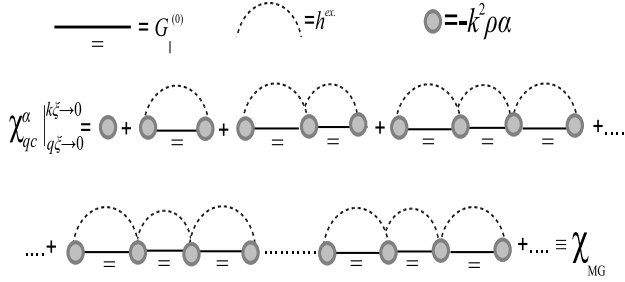}
\caption{Diagrammatic representation
of the 1PI processes which amount to the MG susceptibility. Only the exclusion volume two-point correlation function, $h^{ex.}(r-\xi)$, is relevant and self-correlations are disregarded. In the electrostatic-long-wavelength limit, $k\xi,q\xi\rightarrow0$, the quasicrystalline approximation is exact.}\label{fig3}
\end{figure}

We investigate to which extent the Schwinger vacuum energy of an effective medium is a good approximation to the actual vacuum energy of radiative modes. We will compare their relation up to order $\rho^{3}$. For the reasons given in the previous section recurrent scattering is negligible and the effective susceptibility is well approximated by that of a Maxwell-Garnett (MG) dielectric.
An MG dielectric is characterized by the fact that the only relevant correlation function is that of an exclusion volume, $h^{ex.}(r-\xi)$. In the electrostatic-long-wavelength limit it is proven that the quasicrystalline approximation is exact \cite{PRLvanTigg} [see Fig.\ref{fig3}] and the effective susceptibility is independent of the precise form of $h^{ex.}(r-\xi)$. In particular, we can use the results of the hard-sphere model without loss of generality. According to this, the electrical susceptibility is the sum of a geometrical series of ratio $\chi^{(2)}_{MG}/\rho\alpha=\rho\alpha/3$, in which  only the bare longitudinal propagator and free-space polarizabilities enter. It yields,
\begin{equation}\label{CHIMG}
\chi_{MG}=\frac{\rho\alpha}{1-\chi_{MG}^{(2)}/\rho\alpha}=\frac{\rho\alpha}{1-\rho\alpha/3},
\end{equation}
and from here and neglecting further renormalization on $\alpha$ for the reasons given above, the effective refractive index  is $n\simeq1+\rho\alpha/2+ (\rho\alpha)^{2}/24+..$. It has been already seen that $\mathcal{F}^{V}_{Sch.}$ contains the free space Lamb energy. At $\mathcal{O}(\rho^{2})$, by inserting the series of $n$ in Eq.(\ref{Fschn}) we get,
\begin{equation}\label{Fschn2}
\mathcal{F}_{Sch.}^{V}|^{\mathcal{O}(\rho^{2})}\simeq -\frac{7\hbar}{48\pi^{2}}\Re{\Bigl\{\int_{0}^{\infty}\textrm{d}\omega\:k^{3}(\rho\alpha)^{2}\Bigr\}}.
\end{equation}
Next we turn to the hard-sphere model for the microscopical computation. The $\zeta$-independent terms of $\phi^{(1,0)}_{\alpha,hs}$ in Eq.(\ref{gammatotp}) amount to the energy of radiative propagating modes (rad) within $\mathcal{F}^{V}_{\mathcal{O}(\rho^{2})}|_{rad}$ upon integration in $\omega$. We show in Appendix \ref{appendyo} that, out of them, $-i\frac{k^{3}}{2\pi}\frac{5}{6}\rho\alpha$ comes from the bulk transverse propagator in $2\phi^{(1,0)}_{\perp}$  \footnote{A comment is in order here to amend some erroneous interpretations in \cite{PRAvc}. In the first place the distinction between coherent and incoherent radiation carried out in Sec.IV of \cite{PRAvc} is erroneous. While Eq.\cite{PRAvc}(54) is correct, its equivalence with Eqs.\cite{PRAvc}(42+44,46,48,50) is not. Consequently, the conclusion that only one LFF appears in the expression of the coherent radiation is incorrect. The correct calculations will be published somewhere else \cite{PRA3}.} [Eqs.(\ref{g1perrra},\ref{g1perrrc},\ref{gammatotppe})]. The remaining $-i\frac{k^{3}}{2\pi}\frac{\rho\alpha}{3}$ comes from the bare transverse propagator in the susceptibility function of $\phi^{(1,0)}_{\parallel}$ [Eqs.(\ref{g1perrrc},\ref{gammatotppa})]. Hence, we can write
\begin{equation}\label{FL2prop}
\mathcal{F}^{V}_{\mathcal{O}(\rho^{2})}|_{rad}\simeq-\frac{7\hbar}{48\pi^{2}}\Re{\Bigl\{\int_{0}^{\infty}\textrm{d}\omega\:k^{3}(\rho\alpha)^{2}\Bigr\}},
\end{equation}
and so $\mathcal{F}_{Sch.}^{V}|^{\mathcal{O}(\rho^{2})}\simeq\mathcal{F}^{V}_{\mathcal{O}(\rho^{2})}|_{rad}$. However this equality is accidental. To see this, it is necessary to express $\mathcal{F}^{V}_{avg}$ and $\mathcal{F}_{Sch.}^{V}$ in comparable terms. To this aim we use the expression of Eq.(\ref{manybodyapprox}) for $\mathcal{F}^{V}_{avg}$ but with $\alpha$ instead of $\alpha_{0}$ there, and the extended version of $\mathcal{F}_{Sch.}^{V}$ which incorporates spatial dispersion \cite{Schwinger2105}, $\mathcal{F}_{Sch.}^{V}|^{ext}$. Inserting Eq.(\ref{infunctionofG}) into Eq.(\ref{FSch}) for $\mathcal{F}_{Sch.}^{V}$ but with spatial dispersion we have,
\begin{equation}
\mathcal{F}_{Sch.}^{V}|^{ext}=-2\hbar\Im{}\Bigl\{\int_{0}^{\infty}\frac{\textrm{d}\omega}{2\pi} \int\frac{\textrm{d}^{3}q}{(2\pi)^{3}}\ln{}\Bigl[1-\rho\tilde{\alpha}\sum_{m=0}k^{2}\mathcal{G}^{(m)}_{\perp}(q)\Bigr]\Bigr\},\label{bulko}
\end{equation}
Further, we expand the latter equation up to order $\rho^{3}$ \footnote{The expansion of the logarithms in Eq.(\ref{bulko}) yields a series of $q$-integrals whose integrands are products of powers of polarization propagators of the form $[\mathcal{G}_{\perp,\parallel}^{(m)}]^{s}$. Each factor $\mathcal{G}_{\perp,\parallel}^{(m)}$ is an $m$-scattering loop and the sum of the exponents, $s$, is the loop order --eg. $G_{\perp,\parallel}^{(0)}[\mathcal{G}_{\perp,\parallel}^{(2)}]^{2}\mathcal{G}_{\perp,\parallel}^{(1)}$ is a four-loop term made of one free-space, one single-scattering and two double-scattering loops.},
\begin{eqnarray}
\mathcal{F}_{Sch.}^{V}|^{ext}&\simeq&\mathcal{F}_{V}^{(1)}+\hbar\Im{}\Bigl\{\int_{0}^{\infty}\frac{\textrm{d}\omega}{2\pi}\nonumber\\
&\times&\Bigl[\rho^{2}\Bigl(2\alpha \phi^{(1)}_{\alpha\perp}/\rho+\alpha^{2}k^{4}\int\frac{\textrm{d}^{3}q}{(2\pi)^{3}}[G_{\perp}^{(0)}]^{2}\label{Fbulk2}
\\&+&i\alpha^{2}\frac{k^{3}}{2\pi}\phi^{(1)}_{\alpha}/\rho\Bigr)\nonumber\\
&+&\rho^{3}\Bigl(2\alpha \phi^{(2,0)}_{\alpha\perp}/\rho^{2}+\frac{2}{3}\alpha^{3}k^{6}\int\frac{\textrm{d}^{3}q}{(2\pi)^{3}}[G_{\perp}^{(0)}]^{3}
\nonumber\\&+&2\alpha^{2}k^{4}\int\frac{\textrm{d}^{3}q}{(2\pi)^{3}}G_{\perp}^{(0)}
\mathcal{G}^{(1,0)}_{\alpha\perp}/\rho+\mathcal{O}(\alpha^{4})\Bigr)\Bigr]\Bigr\}.\label{Fbulk3}
\end{eqnarray}
\indent On the other hand, for the evaluation of the radiative modes of $\mathcal{F}^{V}_{avg}$ we use the expressions for the MG $\phi$ factors and effective transverse propagator restricted to radiative modes \cite{PRAvc},
\begin{equation}
\mathcal{G}^{MG}_{\perp}(q)=\mathcal{L}_{LL}G^{eff}_{MG\perp}(q),\quad\phi_{MG}=-i\frac{k^{3}}{2\pi}\mathcal{L}_{LL}^{2}n,\label{phiMG}
\end{equation}
\begin{equation}
2\phi_{MG\perp}=-i\frac{k^{3}}{2\pi}\mathcal{L}_{LL}^{2}n,\nonumber\quad\phi_{MG\parallel}=\phi_{MG}-2\phi_{MG\perp},\nonumber
\end{equation}
with $\mathcal{L}_{LL}=\frac{\chi_{MG}+3}{3}$ and $G^{eff}_{MG\perp}(q)=[(\chi_{MG}+1)k^{2}-q^{2}]^{-1}$.\\
\indent From  Eq.(\ref{Fbulk2}) we have that the first term there contains only the transverse modes in $\mathcal{F}^{V}_{\mathcal{O}(\rho^{2})}$. Its restriction to radiative modes amounts to $-\frac{5\hbar}{24\pi^{2}}\Re{\{\int_{0}^{\infty}\textrm{d}\omega\:k^{3}(\rho\alpha)^{2}\}}$. On the other hand the second term of Eq.(\ref{Fbulk2}) amounts to $\frac{\hbar}{16\pi^{2}}\Re{\{\int_{0}^{\infty}\textrm{d}\omega\:k^{3}(\rho\alpha)^{2}\}}$.
Adding up the last two quantities we obtain Eq.(\ref{Fschn2}) which equals $\mathcal{F}^{V}_{\mathcal{O}(\rho^{2})}|_{rad}$, even though no longitudinal terms enter the Schwinger result. The reason for this equivalence is accidental, since the $\mathcal{O}(\rho^{2})$ radiative term of the difference $\mathcal{F}^{V}_{avg}|^{\perp}-\mathcal{F}_{Sch.}^{V}|^{ext}$ in  Eq.(\ref{discrep})
is equivalent, by reciprocity \cite{PRAvc}, to the longitudinal term in $\mathcal{F}^{V}_{\mathcal{O}(\rho^{2})}|_{rad}$ but with opposite sign.
This relation is not model dependent. Note also that  by incorporating spatial dispersion in $\mathcal{F}^{V}_{Sch.}|^{ext}$, the term $2\phi_{\perp}^{(1,0)}$
contains only one half of the non-propagating term in Eq.(\ref{gammatotp}) proportional to  $\zeta^{-1}_{0}$. The rest of near field terms
in $\mathcal{F}^{V}_{\mathcal{O}(\rho^{2})}$ are disregarded in $\mathcal{F}^{V}_{Sch.}|_{\mathcal{O}(\rho^{2})}^{ext}$.\\
\indent Next we show that the accidental coincidence of $\mathcal{F}_{Sch.}^{V}|_{\mathcal{O}(\rho^{2})}^{ext}$ and $\mathcal{F}^{V}_{\mathcal{O}(\rho^{2})}|_{rad}$ breaks down at $\mathcal{O}(\rho^{3})$. Either by expanding the integrand of Eq.(\ref{Fschn}) or by direct integration of the $\mathcal{O}(\rho^{3})$ terms in Eq.(\ref{Fbulk3}) we get,
\begin{equation}
\mathcal{F}^{V}_{Sch.}|_{\mathcal{O}(\rho^{3})}^{ext}\simeq-\frac{17\hbar}{288\pi^{2}}\Re{\Bigl\{\int_{0}^{\infty}\textrm{d}\omega\:k^{3}(\rho\alpha)^{3}\Bigr\}},
\end{equation}
while the insertion of Eq.(\ref{phiMG}) into Eq.(\ref{manybodyapprox}) yields,
\begin{equation}
\mathcal{F}^{V}_{\mathcal{O}(\rho^{3})}|_{rad}\simeq-\frac{17\hbar}{144\pi^{2}}\Re{\Bigl\{\int_{0}^{\infty}\textrm{d}\omega\:k^{3}(\rho\alpha)^{3}\Bigr\}}
=2\mathcal{F}^{V}_{Sch.}|_{\mathcal{O}(\rho^{3})}^{ext}.
\end{equation}

\section{The Lamb shift from Schwinger's effective medium vacuum energy}\label{discy}

Schaden, Spruch and Zhou (SSZ) \cite{Schaden} have computed the Lamb shift at leading order in $\rho$ from the Schwinger energy of an effective medium in Eq.(\ref{Fschn}), i.e., the free-space Lamb shift. For the reasons explained in Sec.\ref{S4B} Schwinger's approach yields the correct result at leading order for $\chi_{eff}\simeq\rho\alpha_{0}$. Nonetheless, the authors of \cite{Schaden} have applied a more general procedure consisting on computing the Lamb shift out of the variation of $\mathcal{F}^{V}_{Sch.}$ w.r.t. "small" variations of a background refractive index. That is, adopting our nomenclature, they have used the variation of Eq.(\ref{Scheer22224}) together with the result of Eq.(\ref{LSHSCH}) and $\delta\chi_{eff}=2n\Delta n$,
\begin{equation}\label{Schaden}
\mathcal{E}^{LSh}_{SSZ}=\rho^{-1}\Delta_{n}\mathcal{F}_{Sch.}^{V}=\frac{-\hbar}{4\pi^{2}c^{3}\rho}\Re{}\bigl\{\int_{0}^{\infty}\textrm{d}\omega\:\omega^{3}n\Delta\chi_{eff}\bigr\}.
\end{equation}
For a dilute medium, $n\simeq1$, $\Delta\chi_{eff}=\rho\alpha_{0}$, and
\begin{equation}\label{SSZ}
\mathcal{E}^{LSh}_{SSZ}=\frac{-\hbar}{4\pi^{2}c^{3}\rho}\Re{}\bigl\{\int_{0}^{\infty}\textrm{d}\omega\:\omega^{3}\rho\alpha_{0}\bigr\},
\end{equation}
which is nothing but the free-space Lamb shift. This is analogous to the computation carried out by Feynman, Power and Milonni \cite{Feynman,Power,MilonniScripta2} in the limit $\rho\mathcal{V}\rightarrow1$ in which the medium consists of an only dipole in free space--$\mathcal{V}$ being the sample volume. More specifically, the authors of \cite{Schaden} have computed the difference of the free-space Lamb shift between two dielectric states, $I$ and $II$, with  $\Delta\chi_{eff}=\rho(\alpha^{II}_{0}-\alpha_{0}^{I})$.\\
\indent The computation of Milonni, Schaden and Spruch (MSS) in \cite{Milonnietal} combines elements of both \cite{Feynman,Power,MilonniScripta2} and \cite{Schaden}. That is, while in \cite{Milonnietal} the variation of the bulk energy density is calculated between two different states of molecular dielectrics as in \cite{Schaden}, the difference between the states is given by the difference on the polarizability of only one of the dipoles as in \cite{Feynman,Power,MilonniScripta2}. Under the assumption that the dipole in question is randomly placed and the medium behaves as a continuum of refractive index $n\gg\rho\alpha^{I,II}_{0}/2$, the authors take $\Delta \chi_{eff}=\rho(\alpha^{II}_{0}-\alpha^{I}_{0})$ and, using Eq.(\ref{Schaden}), they get
\begin{equation}\label{MSS1}
\mathcal{E}^{LSh}_{MSS}=\frac{-\hbar}{4\pi^{2}c^{3}\rho}\Re{}\bigl\{\int_{0}^{\infty}\textrm{d}\omega\:\omega^{3}
n\rho(\alpha_{0}^{II}-\alpha_{0}^{I})\bigr\}.
\end{equation}
Likewise, for the Lamb shift due to the presence/absence of the background medium (BGM), $\mathcal{E}^{LSh}_{MSS}|^{I-II}_{BGM}=-\frac{\hbar}{4\pi^{2}}\Re{\{\int\textrm{d}\omega(\omega^{3}/c^{3})(n-1)(\alpha^{II}_{0}-\alpha^{I}_{0})\}}$.\\
\indent The authors of \cite{Milonnietal} had already warned that the derivation of the Lamb shift this way might need to be corrected by local field factors in high ordered systems. We have proved in Sec. \ref{DiscLSch} that this is indeed the case under any circumstance, no matter the degree of order. As it was mentioned there, spatial dispersion and longitudinal modes are to be added to Schwinger's formula in order to obtain the correct result. Even if the host medium can be treated as a continuum, the microscopical calculation of the Lamb shift on the simplest geometries for the embedding of a dipole contains LFFs. At leading order, near field and other radiative factors enter the Lamb shift through the terms of $\phi^{(1)}$ in $\mathcal{E}^{LSh}_{\mathcal{O}(\rho)}$. For instance, for a small  Onsager cavity (Ons.), those terms are the ones in Eq.(\ref{gammatotp}) with $\xi$ being the radius of the cavity and $k\xi<1$ \cite{Tomas,Duang06,WelshTomas}. Thus, instead of $\mathcal{E}^{LSh}_{MSS}|^{I-II}_{BGM}$ we obtain at leading order in $(n-1)$ \footnote{The approximate expression of $\phi^{(1,0)}_{\alpha,hs}$ for $\zeta<1$ in Eq.(\ref{gammatotp}) have been used for simplicity since the exponential factor in the integrand of Eq.(\ref{gammatot}) suppresses frequencies greater than $c/(2\xi)$ and the singularities of the integrand locate around $\omega_{0}\ll c/(2\xi)$ . The $\xi$-dependent terms may be more relevant than the radiative ones if the host medium is highly dissipative within the frequency range of integration.},
\begin{eqnarray}
\mathcal{E}^{LSh}_{\mathcal{O}(n-1)}|^{I-II}_{Ons.}&\simeq&\frac{7}{3}\mathcal{E}^{LSh}_{MSS}|^{I-II}_{BGM}\label{uma}\\
&-&\frac{\hbar}{2\pi^{2}}\Im\Bigl{\{\int\textrm{d}\omega(n-1)(\alpha_{0}^{II}-\alpha_{0}^{I})[\frac{1}{\xi^{3}}+\frac{\omega^{2}}{c^{2}\xi}]\Bigr\}}.\nonumber
\end{eqnarray}

\section{Continuum approach to the Binding Energy of an Effective Medium}\label{Continuum}

In this Section we investigate the possibility of quantifying, at least partially, the binding energy through optical observations. This is motivated by the conjecture raised in the introduction on the correspondence between the shift in the spectrum of optical modes and that of energy levels. Against this conjecture we have that there does not exist \emph{a priori} a simple relation between the density of states for emission \cite{PRAvc}, $\mathcal{N}^{emiss}=-2\Im{\{\phi\}}/(\pi\omega)$,
 and that for the vacuum energy, $\mathcal{N}^{V}_{avg}=\frac{-2}{\omega}\int_{0}^{\alpha_{0}}
\delta \alpha_{0}^{'}\frac{\phi}{1+\alpha_{0}^{'}\phi}$. In favor, we have already found that $\mathcal{E}^{LSh}_{avg}$ can be expressed in function of the electrical susceptibility only [Eq.(\ref{eerr22224})]. Moreover, in the quasicrystalline approximation $\mathcal{F}^{V}_{qc}$ is a function of $\chi_{\perp,\parallel}(q)$ [Eqs.(\ref{laFqc}-\ref{Flamb})]. Further, the qc approximation turns into exact in the long-wavelength of the effective medium theory of an MG dielectric. For this reason we will concentrate on the MG model. Nonetheless, we must bear in mind that the MG model neglects recurrent scattering in $\chi_{MG}$, which might be relevant, even in the long-wavelength limit, in dense media for which $\xi^{3}\sim\rho^{-1}$ -- cf. \cite{Felderhof3,Alder1} and Appendix \ref{appendy}.

\subsection{The Lorentz-Lorenz shift}
The problem of the binding energy of an effective medium has been addressed by Bullough and Obada \cite{Obada} in a molecular crystal and by ourselves in \cite{PRAvc}. In the latter reference it has been found that in the long-wavelength limit $\mathcal{F}^{V}_{qc}$ can be split into the energy of transverse bulk modes and that of LFFs and longitudinal bulk modes according to
\begin{equation}\label{decompeff}
\mathcal{F}^{V}_{MG}=\mathcal{F}_{Sch.}^{V}+\Delta\mathcal{F}^{V}_{MG}.
\end{equation}
Here, $\mathcal{F}_{Sch.}^{V}$ is given in Eq.(\ref{Fschn}) and
\begin{equation}
\Delta\mathcal{F}^{V}_{MG}=-\hbar\int_{0}^{\infty}\frac{\textrm{d}\omega}{2\pi}
\int\frac{\textrm{d}^{3}q}{(2\pi)^{3}}\Im{\Bigl\{\ln{}\Bigl[\frac{\chi_{MG}^{3}[\alpha_{0}]}{(\rho\alpha_{0})^{3}\epsilon_{MG}[\alpha_{0}]}\Bigr]\Bigr\}},\label{OBstat}
\end{equation}
where $\chi_{MG}$ are functions of $\alpha_{0}$ in the qc approximation. An identical expression has been reported by Bullough and Obada \cite{Obada}, who have interpreted $\Delta\mathcal{F}^{V}_{MG}$ as the electrostatic binding energy. However, the decomposition in Eq.(\ref{decompeff}), the identification of $\mathcal{F}_{Sch.}^{V}$ with the radiative energy and the identification of $\Delta\mathcal{F}^{V}_{MG}$ with the electrostatic energy are all questionable.\\
\indent Regarding the radiative energy, it was found in Sec.\ref{DiscLSch} that not all the radiative energy is given by $\mathcal{F}_{Sch.}^{V}$. In the first place, it was shown in \cite{PRAvc} that $\mathcal{L}_{\perp}(q)$ induces radiation on the surrounding dipoles around an emitter, acting as a mediator of non-radiative energy transfer. For this to be the case, $\mathcal{L}_{\perp}(q)$ must appear coupled to $G_{\perp}(q)$ prior to integration in $q$. Clearly, the decomposition of Eqs.(\ref{decompeff}) precludes this. On top of that, there is also energy in $\mathcal{F}^{V}_{\parallel,avg}$ carried by indirect radiation modes which are missing in $\mathcal{F}_{Sch.}^{V}$. It comes from the transverse modes within $\chi_{\parallel}(q)$ which are neglected in the long-wavelength limit [eg., the term $-i\frac{k^{3}}{2\pi}\frac{\rho\alpha}{3}$ mentioned before Eq.(\ref{FL2prop}) belongs to indirect radiation].\\
\indent Regarding the electrostatic energy, it is the EM vacuum energy obtained in the electrostatic limit, $c\rightarrow\infty$. Following \cite{Obada}, such a limit must be taken prior to the long-wavelength approximation, making this way radiative modes vanish. On the other hand, for $q\xi\rightarrow0$  the transverse susceptibility equals the longitudinal one because only electrostatic modes enter $\chi_{\perp}(q\xi=0)$. However, as argued in the previous paragraph,   $\mathcal{L}_{\perp}(q\xi=0)$ has a physical meaning when coupled to $G_{\perp}$ acting as an inductor of radiation. Therefore, if radiation is precluded for $c\rightarrow\infty$, $\mathcal{G}_{\perp}^{c\rightarrow\infty}$ does not contribute to the electrostatic energy and neither do the transverse LFFs. As a result, transverse LFFs should not enter the vacuum energy in the electrostatic-long-wavelength approximation despite their presence in Eq.(\ref{OBstat}) for $q\xi\rightarrow0$. Thus, we find that the energy of electrostatic modes in the long-wavelength limit is
\begin{equation}\label{vstat}
\mathcal{F}^{V}_{eff}|_{stat}=-\hbar\int_{0}^{\infty}\frac{\textrm{d}\omega}{2\pi}
\int\frac{\textrm{d}^{3}q}{(2\pi)^{3}}\Im{\Bigl\{\ln{}\Bigl[\frac{\chi_{MG}[\alpha_{0}]}{\epsilon_{MG}[\alpha_{0}]\rho\alpha_{0}}\Bigr]\Bigr\}},
\end{equation}
which equals  $\Delta\mathcal{F}^{V}_{MG}$ but for the absence of the two transverse LFFs.\\
\indent Next, let us make explicit calculations in the MG model with bare polarizabilities. According to it, the electrical susceptibility is the sum of a geometrical series of ratio $\chi^{(2)}_{MG}/\rho\alpha_{0}=\rho\alpha_{0}/3$, in which  only the bare longitudinal propagator enters. Neglecting radiative corrections in Eq.(\ref{CHIMG}) we get,
\begin{equation}\label{chiMG}
\chi_{MG}[\alpha_{0}]=\frac{\rho\frac{2\mu^{2}\omega_{0}}{3\hbar\epsilon_{0}}}{\omega_{0}^{2}
-\omega^{2}-\rho\frac{2\mu^{2}\omega_{0}}{9\hbar\epsilon_{0}}}.
\end{equation}
\indent The question we aim to address is whether the shift of the resonant frequency of $\chi_{MG}$ w.r.t. that of $\alpha_{0}$ has any counterpart in the binding energy of the dipoles in the dielectric. That frequency shift is known as the Lorentz-Lorenz (LL) shift, and it is observable in optics experiments \cite{PRLMakietal}\footnote{Despite of the fact that there are experimental
evidences for $\Delta\omega_{LL}$, it is also known
that the MG model fails to provide accurate results close
the resonance. The reason being that it does not incorporate
the dominant recurrent scattering \cite{Felderhof3}.},
\begin{eqnarray}
\Delta\omega_{LL}&=&\omega_{0}[\sqrt{1-\rho\frac{2\mu^{2}\omega_{0}}{9\hbar\epsilon_{0}\omega_{res}^{2}}}-1]\nonumber\\&=&-\rho\frac{\mu^{2}}{9\hbar\epsilon_{0}}+\mathcal{O}(\rho^{2}).
\end{eqnarray}
This shift is clearly the result of the renormalization of $\chi_{MG}$ in Eq.(\ref{chiMG}) by the electrostatic modes within $\chi_{MG}^{(2)}$. Therefore, its energetic counterpart must find in the formula for $\mathcal{F}^{V}_{eff}|_{stat}$. Since no radiative corrections enter the integrand of Eq.(\ref{vstat}) at all, it does not contain imaginary terms and we can write it as a sum over modes. That is, upon using the regularization $\int\frac{\textrm{d}^{3}q}{(2\pi)^{3}}=\rho$, the $\omega$ integral reduces to the sum of the poles  minus the sum of the zeros of the factors within the logarithm. Using the formulas of an MG dielectric we obtain,
\begin{equation}\label{vstat2}
\mathcal{F}^{V}_{eff}|_{stat}=\rho\frac{-\hbar\omega_{0}}{2}[\sqrt{1+\rho\frac{4\mu^{2}}{9\hbar\epsilon_{0}\omega_{0}}}-1].
\end{equation}
At leading order in $\rho$, $\mathcal{F}^{V}_{eff}|_{stat}\simeq\rho\hbar\Delta\omega_{LL}$ holds.
Because, by assumption, no electrostatic-long-wavelength modes renormalize the single particle polarizability alone, we infer that  $\mathcal{F}^{V}_{eff}|_{stat}$ is the binding energy of collective d.o.f. due to long-range-electrostatic interactions. In a fluorescence experiment $\Delta\omega_{LL}$ would be observed through the emission of several excited atoms of the same cluster decaying at a time.

\subsection{Discussion on the continuum approach}

Our microscopical approach together with the calculations of the previous subsection allow us to answer two of the questions posed by Bullough in \cite{BulloughJPhysA5}. The first one was whether the knowledge of the refractive index spectrum is sufficient to estimate the vacuum energy of long-wavelength fluctuations. Our results mean that neither the radiative nor the electrostatic energy can be correctly accounted for this way. The second question was concerned with the role of the Lorentz field in the continuum approach to the binding energy. We have proved, by strict isolation of the vacuum energy of long-wavelength longitudinal modes, that the Lorentz field gives rise to the LL shift and hence to $\mathcal{F}^{V}_{eff}|_{stat}$.\\
\indent Regarding the radiative energy, we conclude from the previous sections that the knowledge of the spectrum of the refractive index is insufficient
for the quantification of the total radiative energy. In the first place, part of the radiation is not accounted for in the
(bulk) spectrum of $\mathcal{F}_{Sch.}^{V}$. Second, the microscopical approach of Sec.\ref{S5} and Appendix \ref{appendyo} shows that the transverse
propagators within the LFFs terms amount to radiative modes which are disregarded in the long-wavelength limit of Eqs.(\ref{bulky})-(\ref{Flamb}).\\
\indent Regarding the electrostatic energy, we first observe that the relation $\mathcal{F}^{V}_{eff}|_{stat}\simeq\rho\hbar\Delta\omega_{LL}$ must not be interpreted as a quantitative estimate of the binding energy but as a result of consistency. That is, this equation means that if only the longitudinal long-wavelength modes which enter the renormalization of $\chi_{MG}$ are considered in the computation of the binding energy of clusters, the equivalence must hold. On the contrary, it must not be interpreted as a faithful estimation of the electrostatic binding energy.
To see this it suffices to verify from Eq.(\ref{lu}) that $\rho\mathcal{F}^{V}_{\mathcal{O}(\rho^{2})}|^{norec.}_{hs}$
does not contains terms of the order of Eq.(\ref{vstat2}). In particular, all the $\zeta_{0}$-independent terms in that equation have a
radiative origin and are of the order of $\Gamma_{0}/\omega_{0}$ less than that of $\mathcal{F}^{V}_{eff}|_{stat}$.
We conclude that, contrarily to what suggested in \cite{Obada,BulloughJPhysA5}, the continuum approximation is not even sufficient to estimate the orders of magnitude of electrostatic binding energies \footnote{The sum over modes of Eq.(\ref{OBstat}) yields
$-\rho\hbar\omega_{0}[3-2\sqrt{3/(n^{2}_{0}+2)}-n_{0}\sqrt{3/(n^{2}_{0}+2)}]/2$ \cite{Obada} upon use of the regularization $\frac{\int\textrm{d}^{3}q}{(2\pi)^{3}}=\rho$ and $n_{0}=\sqrt{1+\chi_{MG}(\omega=0)}$. Equivalently, for a generic Lorentzian dielectric constant with field strength factor $f$, it yields $\rho f^{2}\hbar\omega_{0}/24$ \cite{PRAvc}.}.
\section{Summary}\label{la6}
We have carried out a microscopical study of the EM vacuum energy of an isotropic and homogeneous molecular dielectric made of a random distribution of two-level atomic dipoles.\\
\indent Before considering statistical averages, we have shown that for a specific configuration $m$, $\mathcal{F}^{V}_{m}$ can be expressed either as a function of the dipole fluctuations [Eq.(\ref{la1})], as a function of the source EM field fluctuations [Eq.(\ref{la2})], or as a combination of both [Eq.(\ref{la3})].\\
\indent When statistical averages are taken, the Lamb shift is a function of the electrical susceptibility only [Eq.(\ref{eerr22224})], $\chi_{\perp,\parallel}$, and hence can be computed out of optical observations. On the contrary, the total vacuum energy is not, and only a cluster expansion is possible [Eq.(\ref{manybodyapproxchi})]. Only in the quasi-crystalline approximation it is possible to give a closed expression for the vacuum energy in terms of $\chi^{qc}_{\perp,\parallel}$ [Eqs.(\ref{bulky}-\ref{Flamb})].\\
\indent Using a hard-sphere model we have discussed to which extent recurrent scattering terms contribute to the total vacuum energy.
Except for the free-space Lamb energy, needed of a UV cut-off, no other divergences either in frequency or momentum space show up in the rest of the vacuum energy. In momentum space, LFFs kill the short distance divergences in the same manner they do in the spectrum of emission \cite{PRAvc,Duang06}. In frequency space, the UV divergences of retarded modes are exponentially suppressed by a natural cut-off at the wavelength of the order of the correlation length. On the contrary, both momentum and frequency divergences show up in the vacuum energy of an effective medium which is a function of the effective bulk propagator only and contains no LFFs. For this reason we interpret that the LFFs  play the role of spectral functions w.r.t. the spectrum of bulk modes in the sense introduced by Ford \cite{Ford}.\\
\indent Related to the last issue, the Schwinger approach to the vacuum energy of an effective medium is in general insufficient to compute the Lamb shift out of variations of $\mathcal{F}_{Sch.}^{V}$, which contradicts the result in \cite{Schaden} --Secs.\ref{DiscLSch} and \ref{discy}. $\mathcal{F}^{V}_{Sch.}$ does not account for the total vacuum energy and contain artificial divergences because of not taking proper account of LFFs. Nevertheless, $\mathcal{F}^{V}_{Sch.}$ is sufficient to study Casimir forces between macroscopic dielectrics \cite{deRaad}.\\
\indent By evaluating the vacuum energy of the electrostatic-long-wavelength modes of a Maxwell-Garnett dielectric we have obtained that  $\mathcal{F}_{eff}^{V}|_{stat}\simeq\rho\hbar\Delta\omega_{LL}$. This relation has been interpreted as a result of the consistency between the spectrum of $\chi_{MG}$ and the energy of the modes involved in the construction of $\chi_{MG}$. It has been concluded that the knowledge of the spectral form of the refractive index is insufficient either to quantify the energy of radiative modes or to estimate the electrostatic binding energy of a molecular dielectric.

\acknowledgments
We acknowledge financial support from the Scholarship Program 'Ciencias de la Naturaleza' of the Ramon Areces Foundation and the FCT-CERN project CERN/FP/116358/2010.
\appendix
\section{Decomposition of $\phi^{(1,0)}$ and computation of $\phi^{(1,0)}_{ovd}$}\label{appendyo}
We write Eq.(\ref{varphi1}) as a spatial space integral in order to show the radiative and electrostatic nature of the propagators involved. To this aim, we make use of the fact that $\bar{G}_{rad.}^{(0)}(r)$ is totally transverse and $\bar{G}_{stat.}^{(0)}(r)$ is totally longitudinal in Fourier space,
\begin{eqnarray}
2\varphi^{(1,0)}_{\alpha\perp}&=&-k^{2}\rho\alpha\textrm{Tr}\Bigl\{\int\textrm{d}^{3}r
[\bar{G}_{rad.}^{(0)}(r)+\bar{G}_{stat.}^{(0)}(r)]\nonumber\\&\cdot&\bar{G}_{rad.}^{(0)}(r)
[1+h(r-\xi)]\Bigr\}=-k^{2}\rho\alpha\int\textrm{d}^{3}r\nonumber\\
&\times&\textrm{Tr}\Bigl\{\bar{G}_{rad.}^{(0)}(r)\cdot\bar{G}_{rad.}^{(0)}(r)\label{g1perrra}\\&+&
\bar{G}_{rad.}^{(0)}(r)\cdot\bar{G}_{rad.}^{(0)}(r)h(r-\xi)\label{g1perrrb}\\&+&
\bar{G}_{stat.}^{(0)}(r)\cdot\bar{G}_{rad.}^{(0)}(r)h(r-\xi)\Bigr\}\label{g1perrrc},
\end{eqnarray}
\begin{eqnarray}
\varphi^{(1,0)}_{\alpha\parallel}&=&-k^{2}\rho\alpha\textrm{Tr}\Bigl\{\int\textrm{d}^{3}r
[\bar{G}_{rad.}^{(0)}(r)+\bar{G}_{stat.}^{(0)}(r)]\nonumber\\&\cdot&\bar{G}_{stat.}^{(0)}(r)
[1+h(r-\xi)]\Bigr\}=-k^{2}\rho\alpha\int\textrm{d}^{3}r\nonumber\\
&\times&\textrm{Tr}\Bigl\{\bar{G}_{stat.}^{(0)}(r)\cdot\bar{G}_{stat.}^{(0)}(r)\label{g1parrra}\\&+&
\bar{G}_{stat.}^{(0)}(r)\cdot\bar{G}_{stat.}^{(0)}(r)h(r-\xi)\label{g1parrrb}\\&+&
\bar{G}_{rad.}^{(0)}(r)\cdot\bar{G}_{stat.}^{(0)}(r)h(r-\xi)\Bigr\}\label{g1parrrc}.
\end{eqnarray}
Note that it is the presence of $h(r-\xi)$ that makes the crossed terms non-vanishing.\\
\indent Next, particularizing to the hard-sphere model with $h(r-\xi)=-\Theta(r-\xi)$,
\begin{eqnarray}
2\varphi^{(1,0)}_{\alpha\perp}|_{hs}&=&\frac{ke^{i\zeta}}{4\pi\zeta^{3}}\rho\alpha
[2-2i\zeta+e^{i\zeta}(-2+4i\zeta+2\zeta^{2}-i\zeta^{3}))]\nonumber\\
&\simeq&\frac{-k}{2\pi}\rho\alpha\Bigl[\frac{1}{2\zeta}+\frac{5}{6}i\Bigr],\quad \zeta<1,\label{gammatotppe}\\
\varphi^{(1,0)}_{\alpha\parallel}|_{hs}&=&\frac{-k}{2\pi}\rho\alpha e^{i\zeta}
\Bigl[\frac{1}{\zeta^{3}}-\frac{i}{\zeta^{2}}\Bigr]\nonumber\\
&\simeq&\frac{-k}{2\pi}\rho\alpha\Bigl[\frac{1}{\zeta^{3}}+\frac{1}{2\zeta}+\frac{i}{3}\Bigr],\quad \zeta<1,\label{gammatotppa}
\end{eqnarray}
where $\zeta=k\xi$. The $\zeta$-independent terms correspond to the long-wavelength propagating modes. In Fourier space, they are given by the poles of the radiative propagators in Eqs.(\ref{g1perrra},\ref{g1perrrc},\ref{g1parrrc}).\\
\indent Next, we estimate the contribution of the two-point overdensity correlation function to $\phi^{(1,0)}$ in $\mathcal{E}^{LSh}_{\mathcal{O}(\rho)},\mathcal{F}^{V}_{\mathcal{O}(\rho^{2})}$. Let  $h^{ovd}(r-\xi)=\xi C\delta^{(1)}(r-\xi)$, where $C$ is a positive constant which accounts for the molecular coordination number. It suffices to integrate in angles to obtain,
\begin{eqnarray}
\phi^{(1,0)}_{ovd}&=&\textrm{Tr}\Bigl\{\int\textrm{d}^{3}r
\:\bar{G}^{(0)}(\vec{r})(-k^{4}\rho\tilde{\alpha})
\bar{G}^{(0)}(\vec{r})\nonumber\\&\times&\xi C\delta^{(1)}(r-\xi)\Bigr\}\nonumber\\
&=&\frac{-k^{3}}{2\pi}C\rho\tilde{\alpha}e^{2i\zeta}
\Bigl[\frac{3}{\zeta^{3}}-\frac{6}{\zeta^{2}}i-\frac{5}{\zeta}+2i+\zeta\Bigr].\label{gammaovd}
\end{eqnarray}

\section{Computation of $\phi^{(1)}_{\alpha}$ and $\bar{\chi}^{(2)}_{\alpha rec}$ including recurrent scattering}\label{appendy}
\begin{figure}[h]
\includegraphics[height=5.5cm,width=8.5cm,clip]{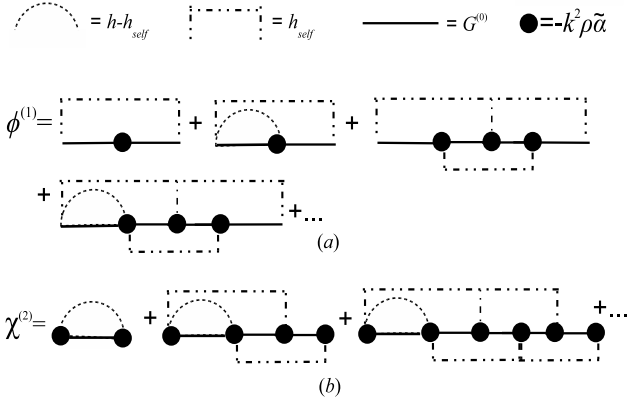}
\caption{Diagrammatic representation
of the series of recurrent scattering diagrams which amount to $(a)$ $\phi^{(1)}$ and $(b)$ $\bar{\chi}^{(2)}$. In the latter series the first diagram corresponds to $\bar{\chi}^{(2,0)}$ while the remaining ones belong to $\bar{\chi}^{(2)}_{rec}$.}\label{fig4}
\end{figure}

We proceed to sum up the infinite series of recurrent scattering diagrams which contribute to $\phi^{(1)}_{\alpha}$ in $\mathcal{E}^{LSh}_{\mathcal{O}(\rho)},\mathcal{F}^{V}_{\mathcal{O}(\rho^{2})}$. The series is pictured diagrammatically in Fig.\ref{fig4}$(a)$. Replacing $\tilde{\alpha}$ with $\alpha$ the sum reads,
\begin{eqnarray}
\phi^{(1)}_{\alpha}&=&-k^{4}\rho\alpha\textrm{Tr}\Bigl\{\int\textrm{d}^{3}r\bar{G}^{(0)}(\vec{r})\label{firec}\\
&\cdot&\sum_{m=0}(k^{2}\alpha)^{2m}
[\bar{G}^{(0)}(\vec{r})]^{2m}\:\cdot\bar{G}^{(0)}(\vec{r})[1+h(r-\xi)]\Bigr\},\nonumber
\end{eqnarray}
where the scatterers are all taken bare to keep the order $\rho$ in the series. An analytical expression for this sum can be given. We follow the computation of \cite{vanTiggrecur} in which the authors decompose $\bar{G}^{(0)}(\vec{r})$ in transverse and longitudinal components w.r.t. the position vector, $\vec{r}$,
\begin{equation}
\bar{G}^{(0)}(\vec{r})=P(r)[\bar{\mathbb{I}}-\hat{r}\otimes\hat{r}]+Q(r)\hat{r}\otimes\hat{r},
\end{equation}
with
\begin{eqnarray}
P(r)&=&\frac{-e^{ikr}}{4\pi r}[1+i/(kr)-1/(kr)^2],\nonumber\\
Q(r)&=&\frac{-e^{ikr}}{4\pi r}[-2i/(kr)+2/(kr)^2].\nonumber
\end{eqnarray}
In terms of $P$, $Q$, Eq.(\ref{firec}) reads,
\begin{eqnarray}
\phi^{(1)}_{\alpha}&=&-k^{4}\rho\alpha\int\textrm{d}^{3}r\Bigr[\frac{2P^{2}}{1-(k^{2}\alpha P)^{2}}\nonumber\\
&+&\frac{Q^{2}}{1-(k^{2}\alpha Q)^{2}}\Bigl][1+h(r-\xi)].\label{A4}
\end{eqnarray}
The integration in $r$ of Eq.(\ref{firec}) and its further integration in $\omega$ in Eq.(\ref{rho2}) require their
expansion in powers of $\alpha$. Hence, the series of Eq.(\ref{expanV}).\\
\indent In a similar fashion, the  full series of recurrent scattering diagrams which amount to $\chi^{(2)}_{rec\perp,\parallel}$ are those on the right-hand side of the equality of Fig.\ref{fig4}$(b)$, but for the first one. Their sum yields,
\begin{eqnarray}\label{chiroc}
\bar{\chi}^{(2)}_{\alpha rec}(\vec{r})&=&-k^{2}(\rho\alpha)^{2}\bar{G}^{(0)}(\vec{r})\label{Xirecs}\\
&\cdot&\sum_{m=1}(-k^{2}\alpha)^{2m}
[\bar{G}^{(0)}(\vec{r})]^{2m}\:[1+h(r-\xi)].\nonumber
\end{eqnarray}
Using the decomposition of $\bar{G}^{(0)}$ in terms of $P$, $Q$, it can be written as
\begin{eqnarray}\label{chirec}
\bar{\chi}^{(2,2)}_{\alpha}(\vec{r})&=&-k^{2}(\rho\alpha)^{2}\Bigr[\frac{P}{1-(k^{2}\alpha P)^{2}}(\bar{\mathbb{I}}-\hat{r}\otimes\hat{r})\nonumber\\&+&\frac{Q}{1-(k^{2}\alpha Q)^{2}}\hat{r}\otimes\hat{r}\Bigl][1+h(r-\xi)].
\end{eqnarray}
At leading order in $(\alpha/\xi^{3})^{2}$, the zero-mode which modifies the original MG formula for the electrical susceptibility is
\begin{equation}
\chi^{(2)}_{\alpha rec}|_{\perp,\parallel}(q\xi=0)\simeq\frac{1}{3}(\rho\tilde{\alpha})^{2}(\tilde{\alpha}/4\pi\xi^{3})^{2}\label{xieffrec}.
\end{equation}

\end{document}